# Molecular-level tuning of cellular autonomy controls the collective behaviors of cell populations

Théo Maire[1,2,3] and Hyun Youk[2,3,*]

[1]Department of Biology, École Normale Supérieure, Paris 75005, France
[2]Department of Bionanoscience,
[3]Kavli Institute of Nanoscience, Delft University of Technology, Delft 2628CJ, the Netherlands
[*]Correspondence: h.youk@tudelft.nl

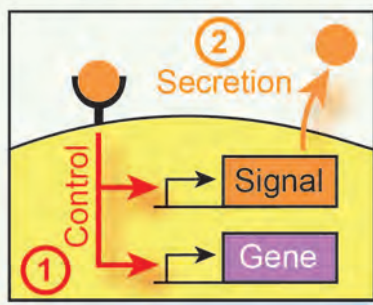
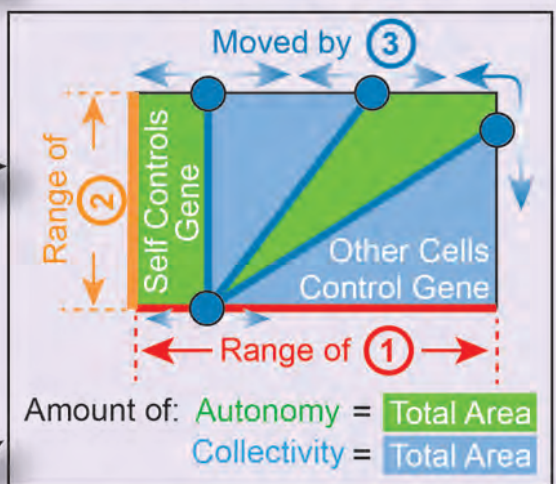
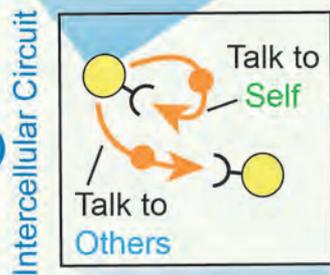
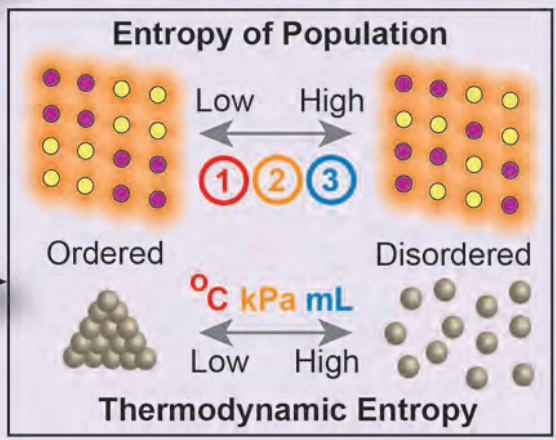
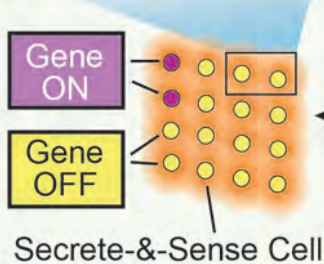

Graphical abstract

**SUMMARY**

**A rigorous understanding of how multicellular behaviors arise from the actions of single cells requires quantitative frameworks that bridge the gap between genetic circuits, the arrangement of cells in space, and population-level behaviors. Here, we provide such a framework for a ubiquitous class of multicellular systems—namely, "secrete-and-sense cells" that communicate by secreting and sensing a signaling molecule. By using formal, mathematical arguments and introducing the concept of a phenotype diagram, we show how these cells tune their degrees of autonomous and collective behavior to realize distinct single-cell and population-level phenotypes; these phenomena have biological analogs, such as quorum sensing or paracrine signaling. We also define the "entropy of population," a measurement of the number of arrangements that a population of cells can assume, and demonstrate how a decrease in the entropy of population accompanies the formation of ordered spatial patterns. Our conceptual framework ties together diverse systems, including tissues and microbes, with common principles.**

**INTRODUCTION**

Intuition tells us that if each cell behaves freely without being influenced by its neighbors, then a population of such autonomous cells would likely behave in a highly uncoordinated manner. On the other hand, if cells strongly influence each other by communicating with one another, then we would expect that a population of such cells would likely behave in a highly coordinated and collective manner. Because removing individual cells' autonomy both shapes the space of possible behaviors that a group of cells can have and limits it, cells likely have more ways to be uncoordinated than to be coordinated with one another. These qualitative and often loosely defined notions about communication among cells are deeply ingrained in our conventional thinking of multicellular behaviors such as the development of embryos, functioning of tissues, and microbes collectively fighting for their survival (Martinez Arias and Stewart, 2003). But many multicellular systems are too complex and involve too many parts (e.g., genetic circuits with many parts, cells at many different locations) for us to use intuition alone to understand and trace the steps that lead to their behaviors (Mehta and Gregor, 2010; Perrimon and Barkai, 2011; Markson and Elowitz, 2014). Casting these loose ideas in a rigorous mathematical framework that connects genetic circuits inside cells to population-level behaviors is crucial for understanding how genetic circuits and cell-cell communication yield multicellular behaviors. Such frameworks would define and quantify the "amount" of cell's freedom, the "amount" of cells' collectiveness, and the potential trade-off between the two. They may also provide common quantitative metrics and concepts that we can apply to many different multicellular systems.

Motivated by these considerations, this paper focuses on how cells use their genetic circuits and cell-cell communication to tune their "degree of autonomy" in order to coordinate their gene expression levels with one another. In particular, we focus on a ubiquitous class of multicellular system: a group of cells that secrete and sense one type of signaling molecule that we call "secrete-and-sense cells" (Figure 1A) (Youk and Lim, 2014). A secrete-and-sense cell can signal to itself ("self-signaling") as well as to other cells ("neighbor-signaling") because it has a receptor that binds the signaling molecule secreted by both itself and its identical neighbors (Figure 1B) (Youk and Lim, 2014; Savir et al., 2012). Secrete-and-sense cells exist in diverse organisms. A special and perhaps the most well known form of secrete-and-sense cells, called "quorum sensing cells", is abundant in the microbial world (Ng and Bassler, 2009). Quorum sensing cells maximize their neighbor-signaling ability while minimizing their self-signaling ability by, for example, having receptors with a very low binding affinity for the signaling molecule. Thus only when there is a sufficiently high density of cells, which results in a high density of the secreted signaling molecule, the cells can capture enough signaling molecules to "turn ON" their genes. Another special form of secrete-and-sense cells, called "autocrine cells", is abundant in the metazoan world (Sporn and Todaro, 1988). Unlike the quorum sensing cells, autocrine cells maximize their self-signaling ability while minimizing their neighbor-signaling ability by, for example, producing large amounts of receptors that bind the signaling molecule. Thus an autocrine cell can easily capture a molecule it had just secreted before the molecule travels far away from the cell. Many microbial and metazoan secrete-and-sense cells, however, have equally dominant self- and neighbor-signaling abilities (Youk and Lim, 2014). Examples of secrete-and-sense cells, each with varying degrees of self- and neighbor-signaling abilities, include the soil amoebae *D.discoideum* that secrete and sense cyclic-AMP to aggregate together (Sgro et al., 2015; Gregor et al., 2010), cells within the embryos of *D. melanogaster* that regulate their fates by secreting and sensing "wingless" (Hooper, 1994), T-cells that secrete and sense IL-2 to regulate their population density (Hart and Alon, 2013; Hart et al., 2014), the marine bacteria *Vibrio harveyi* that quorum sense to collectively generate light (Long et al., 2009), mammary cells whose misregulated secreting-and-sensing of IL-6 is a key step in carcinogenesis (Sansone et al 2007), and *E. coli* cells that use synthetic genetic circuits to quorum sense and form diverse spatial patterns (You et al., 2004; Tanouchi et al. 2008; Song et al. 2009; Pai and You, 2009; Payne et al., 2013). A recent work has revealed that a two-dimensional lattice of hair follicles underneath the skin, despite being macroscopic organs, can also act as point-like secrete-and-sense cells that collectively regenerate hairs (Chen et al., 2015). The ubiquity of secrete-and-sense cells and the fact that despite their diversity, they use common types of genetic circuits to regulate their secretion and sensing (Youk and Lim, 2014), make these cells ideal beds for developing a general theory.

  Here we use a bottom-up approach to derive such a general theory for secrete-and-sense cells. We first show how an isolated secrete-and-sense cell uses its self-signaling (Figure 1B) to regulate its own gene expression. We then show how this cell's autonomous gene regulations (which we call "autonomous behaviors") morph into gene regulations that depend on the neighbors' signaling molecules (which we call "collective behaviors") as we increase the number of neighboring cells and the strength of cell-cell communication. In this process, we define and quantify the cells' degree of autonomy, degree of collectiveness, and the trade-off between the two by representing them as geometric shapes in a "phenotype diagram". We complete our theory by introducing a concept of "entropy of population" that quantifies the consequences of tuning the degree of each cell's autonomy on the whole population. Finally, we give examples of how one can apply our theoretical framework to better understand and engineer secrete-and-sense cells found in nature.

**RESULTS**

**Autonomous behaviors of an isolated secrete-and-sense cell**

We first derive in detail how an isolated secrete-and-sense cell senses its own signaling molecule to regulate its genes. An isolated cell signals only to itself ("self-signaling" in Figure 1B) (Fallon and Lauffenburger, 2000). The concentration of the signaling molecule outside the cell controls the cell's secretion rate of the signaling molecule. Binding of the molecule to the cell's receptor triggers a cascade of molecular events inside the cell (Figure 1C) that either increases (through a positive feedback, Figure 1D) or decreases (through a negative feedback, Figure 1E) the secretion rate by regulating a gene that encodes the signaling molecule (orange box in Figure 1C) (Youk and Lim, 2014). This binding usually also controls one or more "reporter genes" (blue-red box in Figure 1C) that regulate signaling pathways inside the cell (e.g., a master regulator of the stem cell's fate) (Hart et al., 2014; Sgro et al., 2015, Gregor et al., 2010). A sigmoidal function usually describes the cell's secretion rate and the reporter gene's expression level as a function of the signaling molecule's concentration. In many secrete-and-sense cells found in nature, a step-function closely approximates the sigmoidal function (Figures 1D and 1E) (Dayarian et al., 2009; Pai et al., 2014; Hart et al., 2014; Youk and Lim, 2014;

Gregor et al., 2010; Hermsen et al., 2010; Hart et al., 2012). That is, the cell's reporter gene is restricted to be either "ON" or "OFF". An ON cell has a secretion rate $R_{ON}$ and an OFF cell has a secretion rate $R_{OFF}$. $R_{ON}$ is larger than $R_{OFF}$. The cell switches between the two states at a threshold concentration $\widetilde{K}$ (Figures 1D and 1E). The threshold concentration can be tuned, for example, by changing the expression level of the receptor or the receptor's binding affinity for the signaling molecule (Pai and You, 2009; Youk and Lim, 2014). For simplicity, we treat the cell to be point-like. The concentration (denoted $S$) of the signaling molecule with a diffusion constant $D$ and a degradation rate $\gamma$, at a distance $r$ from the cell is governed by the two-dimensional diffusion equation:

$$\frac{\partial S}{\partial t} = \underbrace{D\nabla^2 S}_{\text{diffusion}} - \underbrace{\gamma S}_{\text{degradation}} + \underbrace{R_O \delta(r)}_{\text{secretion}} \quad [1]$$

Here $R_O$ is the secretion rate (equal to either $R_{OFF}$ or $R_{ON}$), and $\delta(r)$ is 1 on the cell ($r = 0$) and zero everywhere else ($r > 0$). The degradation term can represent both a passive degradation of the signaling molecule (i.e., the molecule stochastically degrades) and an active degradation of the molecule by a protease that the cell may secrete at a constant rate. A typical cell repeatedly measures a fluctuating concentration over a long time, averages these multiple measurements, and then uses the average concentration to regulate its genes (Lalanne and François, 2015; Gregor et al., 2007; Govern and ten Wolde, 2012). Since the concentration usually reaches a steady state much faster than the time taken for this averaging, we can focus on how the steady state concentration regulates the cell's behavior. The steady state concentration in two-dimensions forms a gradient that exponentially decays away from the cell:

$$S(r) = S_O \exp\left(\frac{-r}{\lambda}\right) \quad [2]$$

Here $S_O$ is the concentration on the cell's surface. It is proportional to the secretion rate $R_{ON}$ when the cell is ON (then we define $S_O \equiv \tilde{S}_{ON}$) and to $R_{OFF}$ when the cell is OFF (then we define $S_O \equiv \tilde{S}_{OFF}$). $\lambda \equiv \sqrt{D/\gamma}$ is the typical distance that a signaling molecule travels before decaying.

Thus we can consider equation [2] to describe a circular "cloud" of molecules, with radius $\lambda$, centered about the cell (Figure 2A). The cell senses the molecules in this cloud. Here we are assuming that the time taken for the secreted signaling molecules to reach a steady state level (i.e., time taken to build the cloud) is much shorter than the time taken for the cell to determine the concentration and then regulating its genes in response to it. To make meaningful comparisons between the different terms, we divide all concentration terms by the OFF state's concentration $\tilde{S}_{OFF}$:

$$\begin{cases} K = \dfrac{\widetilde{K}}{\tilde{S}_{OFF}} \\ S_{ON} = \dfrac{\tilde{S}_{ON}}{\tilde{S}_{OFF}} \\ \tilde{S}_{OFF} = 1 \end{cases} \quad [3]$$

Thus we now measure all concentrations relative to $\tilde{S}_{OFF}$ (thus $\tilde{S}_{OFF} = 1$). Recast in these rescaled terms, $S_{ON}$ is the concentration on the surface of the ON cell whereas 1 is the concentration on the surface of the OFF cell. From equation [3], we see that $S_{ON}$ and $K$ are the only freely tunable parameters for the cell. Since the cell's state (ON or OFF) depends only on comparing the threshold concentration $K$ with the concentration on the cell surface (Figures 1D and 1E), a function that compares these two values, that we call "phenotype function", would determine what the cell will do next (either maintain or change its current ON/OFF state). Since the concentration on the cell surface is either 1 or $S_{ON}$, we have two phenotype functions, $\varphi_{OFF}$ and $\varphi_{ON}$:

$$\begin{cases} \varphi_{OFF}(K, S_{ON}) = 1 - K \\ \varphi_{ON}(K, S_{ON}) = S_{ON} - K \end{cases} \quad [4]$$

For both the positive and the negative feedbacks, the sign of $\varphi_{OFF}$ determines what the OFF cell will do next (remain OFF or turn ON) while the sign of $\varphi_{ON}$ determines what the ON cell will do next (remain ON or turn OFF). Thus the signs of both functions determine all possible autonomous behaviors ("phenotypes") of the cell. The possible combinations for the signs of $\varphi_{OFF}$ and $\varphi_{ON}$ are:

$$\begin{cases} (1)\ \varphi_{OFF} > 0 \text{ and } \varphi_{ON} > 0 \\ (2)\ \varphi_{OFF} < 0 \text{ and } \varphi_{ON} > 0 \\ (3)\ \varphi_{OFF} < 0 \text{ and } \varphi_{ON} < 0 \end{cases} \quad [5]$$

The scenario in which $\varphi_{OFF} > 0$ and $\varphi_{ON} < 0$ cannot occur because the secretion rate of the ON cell ($R_{ON}$) is larger than the secretion rate of the OFF cell ($R_{OFF}$). Thus the concentration on the surface of the ON cell ($S_{ON}$) is larger than that of the OFF cell ($\tilde{S}_{OFF} = 1$). Thus $\varphi_{ON} > \varphi_{OFF}$ and hence we cannot simultaneously have $\varphi_{OFF} > 0$ and $\varphi_{ON} < 0$. For both the positive and negative feedback regulation, the above three conditions split the plane spanned by $K$ and $S_{ON}$ into three regions (Figures 2B and 2C). Each region represents a distinct phenotype of the cell. Thus we call the resulting two diagrams, one for the positive feedback (Figure 2B) and the other for the negative feedback (Figure 2C), "phenotype diagrams". We deduce the phenotypes represented by each region from the input-output step functions (Figures 1D and 1E). A cell with the positive feedback and a cell with the negative feedback have two phenotypes in common. First the cell turns itself ON and stays ON due to self-signaling ("ON" region in Figures 2B and 2C). Second, the cell's self-signal is insufficient to maintain itself ON so the cell remains OFF ("OFF" region in Figures 2B and 2C). In addition, the positive feedback enables a "bistable" phenotype ("ON & OFF" region in Figure 2B) in which the cell can either stay ON or stay OFF, depending on its past history. The bistable cell can switch between ON and OFF due to external perturbations and stochastic silencing or activation of its secretion. In the case of the negative feedback, the cell can flip back and forth between being ON and OFF over time. This occurs only if the molecule degrades sufficiently fast and its concentration reaches the steady state much faster than the cell can toggle between ON and OFF. The phenotype diagrams (Figures 2B and 2C) are geometric blueprints that tell us how the cell should tune the key parameters, $K$ and $S_{ON}$, to realize these distinct phenotypes.

If the cell were a three-dimensional sphere of radius $R$ instead of being a point (Figure 2D), we would need to solve the three-dimensional diffusion equation instead of equation [1] to obtain the steady state concentration around the cell in three-dimensions. We have performed this calculation (see Supplemental Theoretical Procedures) and found that the cell's radius $R$ does not affect the ratio of $\tilde{S}_{ON}$ to $\tilde{S}_{OFF}$ (Figure S1). Thus if we measure the concentration in units of $\tilde{S}_{OFF}$ (i.e., $\tilde{S}_{OFF} = 1$) through equation [3], then $S_{ON}$ is independent of how big the spherical cell is. As a result, we obtain phenotype diagrams for a spherical cell (Figures 2E and 2F) that are identical to the phenotype diagrams of the point-like cell.

**Entangled web of cell-cell communications in a population**

We now present a general formalism to study a population with an arbitrary number of cells. We first define a "basic unit" (Figure 3A), which serves as our elementary building block of larger populations. It consists of identical secrete-and-sense cells at each corner of a hexagon with an edge length $a_O$. It also has a cell at its center (Figure 3A). To build a population of $N$ cells, we repeatedly tile the basic unit next to each other (Figure 3A) (Our framework is applicable to any polygon besides the hexagon). Our main idea is to pick any arbitrary cell in the population, call it "cell-I" ("I" for Individual), and then analyze how its state (ON or OFF) changes as we tune its communication with all the other cells. We number all the other cells (the "neighbors"), from 1 to $N$-1. The concentration $S_I$ of the signaling molecule sensed by cell-I is the sum of the concentration of the molecule secreted by cell-I (denoted $S_{self}$) (equation [2]) and the concentration of the molecule secreted by all the other cells (denoted $S_{neighbors}$) (Figure 3B):

$$S_I = \underbrace{S_O}_{due\ to\ self\ (=S_{self})} + \underbrace{\sum_{j=1}^{N-1} S_{Oj} exp\left(\frac{-r_j}{L}\right)}_{due\ to\ neighbors\ (=S_{neighbors})} \quad [6]$$

Here $r_j$ is the distance between a $j^{th}$ cell and cell-I in units of the edge length $a_O$. $L \equiv \lambda/a_O$ is the "signaling length", which is the radius of the diffusive signal-cloud (Figure 2A) in units of the edge length. The terms $S_O$ and $S_{Oj}$ depend on the state of cell-I and the $j^{th}$ cell respectively (i.e., they are either $S_{ON}$ or 1).

To compute the concentration $S_I$ sensed by cell-I, we need a system for keeping track of the state of every cell in the population. We let $C$ represent cell-I's state (Figure 3C). $C=1$ denotes an ON cell-I whereas $C=0$ denotes an OFF cell-I. Similarly we let $C_j$ denote the state of the $j^{th}$ neighbor (Figure 3C). Then the string $\Omega = (C_1, C_2,...,C_{N-1})$, which we call "neighbor state", denotes the state of all the neighbors. Moreover the string of $N$ binary digits $(C, \Omega)$, which we call "population state", represents the state of the whole population. Since there are $2^N$ different population states, the concentration $S_I$ has $2^N$ possible values (one for each possible value of $(C, \Omega)$). This is a large number even for a small population size (e.g., for a population of $N=20$ cells, $2^N$ is approximately 1 million). Our challenge then is to reduce this complexity, provide a rigorous description of cell-I's degree of autonomy, and reveal all possible behaviors of the population.

**Phenotype functions for populations**

If we know cell-I's behavior in each neighbor state, then we know how cell-I would behave under all possible neighbor states. First, we deduce cell-I's phenotypes for a fixed state of the neighbors (i.e., fix a value for $\Omega$). For this neighbor state, we define a phenotype function: $\varphi_{C,\Omega}(K, S_{ON}, L) \equiv S_I - K$. To construct cell-I's phenotype diagram for this particular neighbor state, let us first fix the value of signaling length $L$ so that we only need to consider how the values of $(K, S_{ON})$ affect the phenotype function. We note that the values of $(K, S_{ON})$ for which $\varphi_{C,\Omega} = 0$ form a straight line (Figure 3D). We call the region above this line an "activation region" (Figure 3D: green region). In this region, cell-I turns ON because it senses a concentration $S_I$ that is above the threshold concentration $K$ (i.e., $\varphi_{C,\Omega} > 0$). Below the line is "deactivation region" (Figure 3D: brown region). In this region, cell-I turns OFF because it senses a concentration $S_I$ that is below the threshold concentration $K$ (i.e., $\varphi_{C,\Omega} < 0$). Repeating this procedure for every neighbor state in a population of $N$ cells, we would obtain $2^N$ activation regions and deactivation regions. When we overlay all these regions onto one plane, we would obtain a full phenotype diagram that shows all possible behaviors of cell-I because it takes into account every possible state of the neighbors.

**Main design principle: Self-signaling competes with neighbor-signaling to control the cell's autonomy**

We have now established our formalism. But before applying it to a population of an arbitrary size, we now explain the main principle that gives rise to different phenotypes. Our idea is to compare the influence on cell-I by self-signaling with the neighbors' influence. First note that the neighbors have minimal influence on cell-I if all of them are OFF. This minimum concentration that the neighbors can create on cell-I is (by setting $S_{Oj}$=1 for all neighbors in equation [6]):

$$f_N(L) = \sum_{j=1}^{N-1} exp(-r_j/L) \quad\quad [7]$$

For reasons we will see shortly, we call $f_N(L)$ the "signaling strength" function. The maximum concentration that the neighbors can generate is $S_{ON}f_N(L)$, which is realized when all the neighbors are ON. The difference between the maximum and the minimum (denoted $\Delta S_{neighbours}$) represents the range of influence that the neighbors have on cell-I. Analogously the difference between the maximum ($S_{ON}$) and the minimum concentration (1) generated by cell-I on itself (denoted $\Delta S_{self}$) represents the range of influence that self-signaling has on cell-I. Specifically, having $\Delta S_{self}$ larger than $\Delta S_{neighbours}$ ($\Delta S_{self} > \Delta S_{neighbours}$) means that cell-I can sense more of its own signaling molecules than the neighbors' signaling molecules, just as an autocrine cell would. In this case, we find that the signaling strength $f_N(L)$ is less than 1 (Figure 3E). On the other hand, having $\Delta S_{self}$ smaller than $\Delta S_{neighbours}$ ($\Delta S_{self} < \Delta S_{neighbours}$) means that cell-I can sense more signals from its neighbors than from itself, just as a quorum-sensing cell would. In this case, we find that the signaling strength $f_N(L)$ is larger than 1 (Figure 3E). The two cases are separated by a "critical signaling length" $L_c$, whereby the influence of self and neighbors are exactly balanced (i.e., $f_N(L_c)$ = 1).

To state in another way, self-signaling (thus autonomy) dominates when $L$ is less than $L_c$, but signaling between cells (thus collectiveness) dominates when $L$ is larger than $L_c$ (Figure 3E). The critical signaling length $L_c$ depends on the number of cells in the population. Crucially, we can always find a critical length for a population with any number of cells. This means that no matter how many cells form a population, cells can always adjust their signaling length $L$ so that each cell has some degree of autonomy. From here on, we will focus on cells with the positive feedback and not repeat our calculations for cells with the negative feedback because both regulations use our theoretical formalism in the same way.

**Application of our general formalism to a small population: A basic population unit**

We now apply our formalism to a small population - the hexagonal basic unit (Figure 3A). We choose cell-I to be at the center of the hexagon and consider a scenario in which the signaling length $L$ is shorter than the critical length $L_c$. Applying our formalism (see "Theoretical Methods" section), we obtain a phenotype diagram with geometric regions that mark different phenotypes of cell-I (Figure 4A: right panel). It has three types of regions: Activation regions, deactivation regions, and an autonomous bistable region.

The activation regions consist of several sub-regions. One is the autonomous "ON" region in which cell-I autonomously turns itself ON (Figure 4A: orange "ON" region). The others are neighbor-induced activation regions (Figure 4A: green regions denoted $\widetilde{An}$), in which cell-I turns ON only if there are at least *n* ON neighbors (Figure 4B).

The deactivation regions consist of several sub-regions as well. One is the autonomous "OFF" region in which cell-I turns itself OFF through self-signaling. The others are neighbor-induced deactivation regions (Figure 4A: brown regions denoted $\widetilde{Dn}$), in which cell-I turns OFF unless there is more than *n* ON neighbors (Figure 4B).

The autonomous bistable region (Figure 4A: yellow region denoted "ON & OFF") represents the bistable "ON & OFF" phenotype that we previously described for the isolated cell. Here the cell is free to choose between being ON or OFF and is unable to "listen" to its neighbors.

Comparing the phenotype diagram of the isolated cell (Figure 4A: left panel) with that of the basic population unit (Figure 4A: right panel), we see that the global effect of cell-cell signaling is reducing the combined area of the three autonomy regions (Figure 4A: blue, yellow, orange regions) to make room for the neighbor-induced activation regions (the $\widetilde{An}$'s in Figure 4A) and neighbor-induced deactivation regions (the $\widetilde{Dn}$'s in Figure 4A). Despite its reduction, the total area of the autonomy regions remains non-zero, meaning that cell-I can regulate its genes autonomously. Our analysis here shows that the combined area of the autonomy regions is a sensible and a quantitative representation of cells' degree of autonomy. The combined area of the regions representing neighbor-induced phenotypes quantifies the cells' degree of collectiveness.

If the basic unit consists of spherical cells of radius *R,* we obtain a phenotype diagram for the basic unit (Figure S1) that is essentially identical to that of the basic unit composed of point-like cells. The reason is that $S_{ON}$ is independent of *R* if we measure all concentrations relative to $\tilde{S}_{OFF}$ (i.e., $\tilde{S}_{OFF} = 1$) as in the case of an isolated spherical cell (see Supplemental Theoretical Procedures).

**Application of our general formalism: Population of an arbitrary size.**

We now apply our formalism to the most general case: A population with $N$ cells with a positive feedback. Thus we can now allow populations to be of an arbitrarily large size. Applying our formalism (see "Theoretical Methods"), we obtain a phenotype diagram with distinct regions (Figure 4C) whose areas depend on the signaling strength $f_N(L)$.

When the signaling strength is very weak (i.e., $f_N(L)<<1$), there is a finite but nearly negligible signals from the neighbors. Thus we obtain a phenotype diagram (Figure 4C: left panel) that is similar to that of the isolated cell (Figure 4A: left panel). The only difference is that the weak signals from the neighbors have reduced the area of the autonomous bistable region (Figure 4C left panel: yellow "ON & OFF" region) and the area of the autonomous "OFF" region (Figure 4C left panel: blue "OFF" region). This contraction in the areas of the two regions makes room for two new regions: a neighbor-induced activation region (Figure 4C left panel: green region) and a neighbor-induced deactivation region (Figure 4C left panel: brown region). As we did in the case of the basic population unit, we see a decrease in each cell's degree of autonomy (i.e., decrease in combined areas of orange, yellow, and blue regions) and as a trade-off, a corresponding increase in the cells' degree of collectiveness (areas of the green and brown regions).

If we now increase the signaling length $L$ but still keep it below the critical signaling length $L_c$ (Figure 4C: middle panel), the neighbor-induced activation region further expands into and overtakes the autonomous bistable region (Figure 4C middle panel: green invades into yellow). In addition, the neighbor-induced deactivation region further expands into and overtakes the autonomous "OFF" region (Figure 4C middle panel: brown invades into blue). This further increases the cells' degree of collectiveness at the expense of the decrease in the degree of autonomy in the corresponding amount.

If we further increase the signaling length $L$, this time above the critical signaling length $L_c$ (Figure 4C: right panel), the autonomous bistable region vanishes because the neighbor-induced activation region completely overtakes it. The neighbor-induced activation region also invades into the neighbor-induced deactivation region (i.e., green invades into brown region). Their merging results in the creation of a new phenotype region that we call "activation-deactivation region" (Figure 4C right panel: white region). In this region, the neighbors collectively decide whether to activate or deactivate cell-I depending on which of the two is larger: the density of ON neighbors (leads to activation) or the density of OFF neighbors (leads to deactivation). Thus we can think of this region as representing a multicellular bistable switch - a type of quorum sensing (Ng and Bassler, 2009; Pai et al. 2012) that measures the density of ON/OFF cells and their local spatial distributions. It is the multicellular analogue (i.e., dependent on neighbors) of the autonomous bistable switch (Figure 4A: yellow "ON & OFF" region). We will see additional reasons later for why this reasoning makes sense when we analyze population-level dynamics enabled by the activation-deactivation region.

We note that while the cells can increase their signaling length $L$ above the critical length $L_c$ to eliminate their autonomous bistable region (Figure 4C: yellow region), the autonomous ON region (Figure 4C: orange region) and the autonomous OFF region (Figure 4C: blue region) still remain, but the cells in these two regions must solely remain ON or remain OFF, respectively. However, the cells in the autonomous bistable region may "choose": Either stay ON or stay OFF. Increasing the signaling length $L$ gradually eliminates this "freedom" by making the autonomous bistable region vanish. Thus while the cells' degree of autonomy remains non-zero when the signaling strength is above 1 (i.e., $f(L_c) > 1$), the cells' degree of autonomous "choice" (area of the yellow region) completely vanishes.

If we have a population of $N$ spherical cells, we can still apply the formalism that we applied to the population of point-like cells. In fact, our calculations show that the phenotype diagrams for a population of $N$ spherical cells are essentially identical to those of a population of $N$ point-like cells (see Supplemental Theoretical Procedures). There are quantitative differences between the population of point-like and population of spherical cells. Namely, the radius $R$ of the spherical cells affects the signaling strength function (denoted $f_{N,R}(L)$) and the concentration $S_I$ sensed by cell-I (Figure S1). But the signaling strength $f_{N,R}(L)$ of the spherical cells is still divided into the same three regimes (Figure 3E) as the point-like cells.

**Entropy of population connects unicellular freedom with population-level freedom.**

We now ask how the different unicellular phenotypes (Figure 4C) generate population-level dynamics (i.e. connecting middle panel to right panel in Figure 1C). To address this question, we first asked if there are spatial arrangements of ON and OFF cells in which no cell's state (i.e., ON or OFF) would change over time. We say that such a population is in an equilibrium configuration. To search for such equilibrium configurations, we performed computer simulations in which we started with a randomly chosen initial arrangement of ON and OFF cells in a population (see Supplemental Theoretical Procedures). We then computed the concentration $S_I$ for each cell (equation [6]). Then we checked if any cell's state (ON or OFF) changed. If none of the cells' states changed, the initial population is in equilibrium. By repeating this process many times, each time with a different configuration of the population, we counted the number of equilibrium configurations that $N$ cells can form with a particular value of ($K$, $S_{ON}$, $L$). We have done this for a wide range of values of ($K$, $S_{ON}$, $L$). To complement our simulations, we derived an analytical formula that estimates the number of equilibrium populations (denoted $\Omega_E$) for each value of ($K$, $S_{ON}$, $L$) (see Supplemental Theoretical Procedures). To meaningfully interpret and compare the $\Omega_E$ obtained by the two methods, we define "entropy of population":

$$\sigma(K, S_{ON}, L) = \frac{\Omega_E}{2^N} \qquad [8]$$

To see what this represents, note that $2^N$ is the total number of possible population states with $N$ cells (Figure 3C). Thus $\sigma = 1$ represents a maximal population-level disorder (population can be in any configuration) and maximal population-level freedom (any configuration is in equilibrium) while $\sigma=1/2^N$ represents a minimal population-level disorder (everyone is in the same state) and minimal population-level freedom (only one configuration is in equilibrium). The entropy of population is thus a macroscopic (population-level) metric based on the microscopic (unicellular) parameters ($K$, $S_{ON}$, $L$) that measures the number of ways that stable gene expression levels (ON or OFF) can be assigned to cells at different locations. We found that the entropy of population determined by our simulations and formula closely agreed with each other for a wide range of values of ($K$, $S_{ON}$, $L$) (Figures 5A and 5B). We found that the entropy of population decreases when the cell-cell interaction strength $f_N(L)$ increases because cell-cell signaling to increases the cells' coordination (compare top and bottom panels in Figure 5B). We also see that the entropy of population is highest when the cells are in the autonomous bistable region (Figure 5B: yellow "ON & OFF" region). This makes sense because when every cell is completely free to choose its state, the whole population can have the maximal number of possible configurations. The entropy of population thus rigorously captures our qualitative notions about how unicellular autonomy is linked to cell-cell coordination at the population-level.

**Population-level dynamics: Self-organization of spatially ordered patterns from spatially disordered populations**

So far we have determined how a cell can dynamically change its state in response to signals from self and neighbors, and the number of ways that populations can be in equilibrium. The final step of our bottom-up program (Figure 1C) is a determination of how cells within a population reach an equilibrium configuration. Population configurations that are not in equilibrium must, by definition, use cell-cell signaling to readjust the behavior of individual cells until the population reaches one of the equilibrium configurations. Development of spatial patterns, such as stripes and islands, occurs in real and quasi two-dimensional systems such as tissues and embryos (Turing, 1952; Gregor et al., 2005; Ben-Zvi et al., 2008). The general principles that govern how these spatial patterns form from secrete-and-sense cells have been elusive. To gain insights, we investigated if ON and OFF cells that are randomly distributed over space can dynamically self-organize into a population with defined spatial patterns. To quantify the spatial ordering of cells, we define a "clustering index" $I_M$, motivated by a statistical metric called "Moran's $I$" (Moran, 1950; see "Theoretical Methods" section). Our clustering index $I_M$ quantifies how closely ON cells (and thus OFF cells) cluster together in space. The clustering index can be between 0 (spatially disordered state) and 1 (spatially ordered state) (Figure S2). As the clustering index approaches zero, ON and OFF cells become more randomly dispersed in space. As the clustering index approaches one, ON cells become more clustered together in one spatial region (e.g., island of ON cells surrounded by a sea of OFF cells).

For each region of the phenotype diagrams (Figure 4*C*), we used two types of simulations to determine how an initially randomly distributed cells' clustering index (i.e., $I_M=0$) evolved over time (Figure S3). One type of simulation was a "deterministic simulation" in which each cell exactly sensed the concentration of the signaling molecule without making errors (Figure 5C). Another type of simulation was a "stochastic simulation" in which the cells made errors in sensing the concentration of the signaling molecule (Figures S4 and S5). Both types of simulations are similar in spirit to the cellular automata and Ising-type models that researchers have previously used in studying pattern formation in developmental and neuronal systems (Ermentrout and Edelstein-Keshet, 1993; Hopfield, 1982). In both types of simulations, we discovered that if nearly 50% of the cells are initially ON and they are in the "activation-deactivation" region (Figure 5C: white region in phenotype diagram), then a spatially disordered population of cells (i.e., $I_M \sim 0$) has a higher chance of evolving into a population with spatially ordered patterns (i.e., $I_M$ closest to 1) than if the cells were in the activation region or the deactivation region (Figure 5C: compare the three graphs of $I_M$) (also see Figures S4-S6). Intuitively this occurs because for a spatially disordered population to be spatially ordered, the randomly scattered OFF cells and ON cells need to expand or contract their territories to form consolidated islands of OFF and ON cells respectively. The expansion of OFF (and ON) cells requires deactivation (and activation), which enables a clustered region of OFF (and ON) cells to cooperatively create more OFF (and ON) cells in their adjacent regions. Such dynamic regulations of the shape and size of the OFF and ON regions are required to form islands of highly clustered OFF and ON cells. Thus when the activation and deactivation co-exist, both ON and OFF cells can simultaneously regulate their shapes and sizes. This enables a spatially disordered population to evolve into a population with a higher spatial order, more so than when activation alone or deactivation alone is present.

We also observed in our simulations that some spatially disordered populations could maintain their fraction of ON cells at a nearly constant value over time while sharply increasing their spatial ordering (i.e., increasing the $I_M$ to a high value near 1). This resulted in highly defined and striking spatial patterns (highly ordered stripes and islands of ON cells) that are stable for long periods of time (Figure S6). The ordered spatial patterns formed if the cells were in the activation-deactivation region of the phenotype diagram. The entropy of population forms a landscape as a function of the threshold concentration $K$ and the maximal concentration $S_{ON}$ (height of the landscape is represented as a heat map in Figure 5B). This landscape has a minimal basin (i.e., a region of local minimum for the entropy of population) within the activation-deactivation region (Figure 5B: lower panel). In our simulations, we found that cells in this region of minimal entropy formed the most stable and ordered spatial patterns (Figure S6). Moreover, we observed that spatial clustering of cells strongly influence how the ON/OFF state of each cell in a population changes over time (See Supplemental Theoretical Procedures and Figure S7).

In summary, our results show that our quantification of degrees of autonomy and of collectiveness are meaningful in making sense of population-level dynamics, including genetically identical cells self-organizing into defined spatial patterns of the types that we encounter in animal development. In particular, our results reveal that a decrease in the entropy of population accompanied by a strong signaling strength, which creates the activation-deactivation region, is correlated with the cells forming highly ordered spatial patterns (Figure 5C).

**DISCUSSION**

On a conceptual level, we have shown that the cells' degrees of autonomy and of collectiveness - two concepts that are central to all multicellular behaviors that are typically loosely and qualitatively treated - can be sensibly defined, quantified, and tuned. This has practical implications. For example, the gain of autonomy by a few renegade secrete-and-sense cells in a healthy tissue often marks the beginnings of a tumour growth (e.g., renegade secreting-and-sensing of IL-6 by a few cells trigger breast carcinoma) (Sansone et al, 2007; Sporn and Todaro, 1980). Thus quantifying an increase in the autonomy and the decrease in the collectiveness of cells may provide quantitative insights into how tumours arise. Our theory may also aid in quantitatively analyzing how maintaining collectiveness keeps tissues healthy.

On a practical level, our work identified the interconnected relationships among the components of genetic circuits and cell-cell signaling that experimentalists can tune to control the cells' autonomous and collective behaviors. We also identified what these behaviors are. The behaviors can be any features of cells that our idealized ON/OFF genes influence downstream. Cells can tune their threshold concentration, for example, by changing the production level of a transcription factor that mediates the positive or negative feedback (Youk and Lim, 2014) or by changing the abundance of the receptors that bind the signaling molecule (e.g., EGF-receptor in EGF-signaling) (DeWitt et al., 2001). Cells can tune their signaling length, for example, by secreting a protease that degrades the signaling molecule (e.g., Bar1 in budding yeast (Rappaport and Barkai, 2012; Diener et al., 2014), phosphodiesterase in the soil amoebae *D. discoideum* (Gregor et al., 2010)). Our work shows that varying the geometric shape of tissues or organs composed of secreting-and-sensing cells can also tune their signaling length. Researchers have experimentally shown many other ways of tuning these elements (Hart et al., 2014; Sgro et al., 2015; Gregor et al., 2010). Thus our theory provides a readily applicable and common framework for understanding and engineering diverse multicellular systems composed of secrete-and-sense cells. Our results for the cells with binary gene regulation (Figures 1D and 1E) also apply to cells that have a finite Hill coefficient controlling their positive or negative feedbacks (see Supplemental Theoretical Procedures and Figures S8-S10).

Our work also suggests the underappreciated ability of secrete-and-sense cells to generate defined spatial patterns, akin to those seen in development of animals such as the fruit fly (e.g., via secreting-and-sensing Wingless (Hooper, 1994)). Specifically, our work shows that given an initial arrangement of ON and OFF secrete-and-sense cells that is spatially disordered, it is possible for highly ordered spatial patterns such as stripes and islands of ON/OFF cells to emerge, with the caveat that the exact location of the spatial patterns in the field of cells is determined by the initial locations of the ON and OFF cells. Thus, if another mechanism sets up a particular initial pattern, which can be spatially disordered (i.e., $I_M \sim 0$), cell-cell communication among the secrete-and-sense cells can take over and generate highly ordered spatial patterns. This may suggest that tissues and embryos composed of secrete-and-sense cells are ideal candidates for realizing the "Turing-like" patterning mechanism (Turing, 1952). Despite decades of search for multicellular systems that use a patterning mechanism similar to the one proposed by Turing, it has been difficult to conclusively prove in many systems that the observed spatial patterns originate from Turing's mechanism (Economou et al., 2012). The main difficulty has been that Turing's formulation of spatial patterning (Turing, 1952) involves only molecules (activator and inhibitor) but not cells. We suggest that it might be fruitful to investigate how secrete-and-sense cells in the activation-deactivation region of the phenotype diagram (Figure 4C), despite not satisfying exactly the conditions of Turing's activator and inhibitor molecules, may act a cellular analogues of Turing's activators and inhibitors.

We also note that, the entropy of population $\sigma$ describes how many spatial patterns can be stably sustained in a population, and can be rigorously defined even if the only information we have about the population are the values of the three molecular parameters, $S_{ON}$, $K$, and $L$ without knowing anything else. Without knowing anything about the initial ON/OFF state of every or even any cell in the population, the entropy of population will predict precisely how many spatial patterns can arise in the population and how likely it is that these patterns are spatially ordered (through the relationship between $\sigma$ and the spatial clustering index $I_M$). Being able to predict a population-level property without having detailed information about the state of any individual cell makes the entropy of population similar in spirit to the thermodynamic entropy (Landau and Lifshitz, 1980) and the Shannon's informational entropy (Shannon, 1948), both of which quantify a systems-level property without having information about the detailed microstate of the system. Thus the entropy of population allows one to predict how likely the expression level of a gene (e.g., ON/OFF) in each cell in a population would form a spatially ordered pattern, in cases where we cannot experimentally measure the expression levels of a gene in any cell in multicellular systems such as a tissue or a biofilm. This connection between the entropy of population and spatial order is reminiscent of the link between the thermodynamic entropy and the amount of disorder in a physical system, and also of the link between randomness of information in a message and the Shannon informational entropy. It may be fruitful to investigate if there are deeper connections between Shannon's entropy and the entropy of population, given that both deals with how much information is accessible to an experimentalist about a particular system.

    We hope that our work will motivate future studies that use first principles to link genetic circuits with multicellular behaviors. Future works that explore alternative ways of defining and quantifying degrees of autonomy and collectiveness in other types of cells will, together with our theory, provide a rigorous framework for understanding and manipulating multicellular systems.

As we have done here, such studies will reveal how quantitative principles of macroscopic living systems emerge from the microscopic laws of molecular and cellular interactions (Phillips, 2015; Mehta and Gregor, 2010; Perrimon and Barkai, 2011).

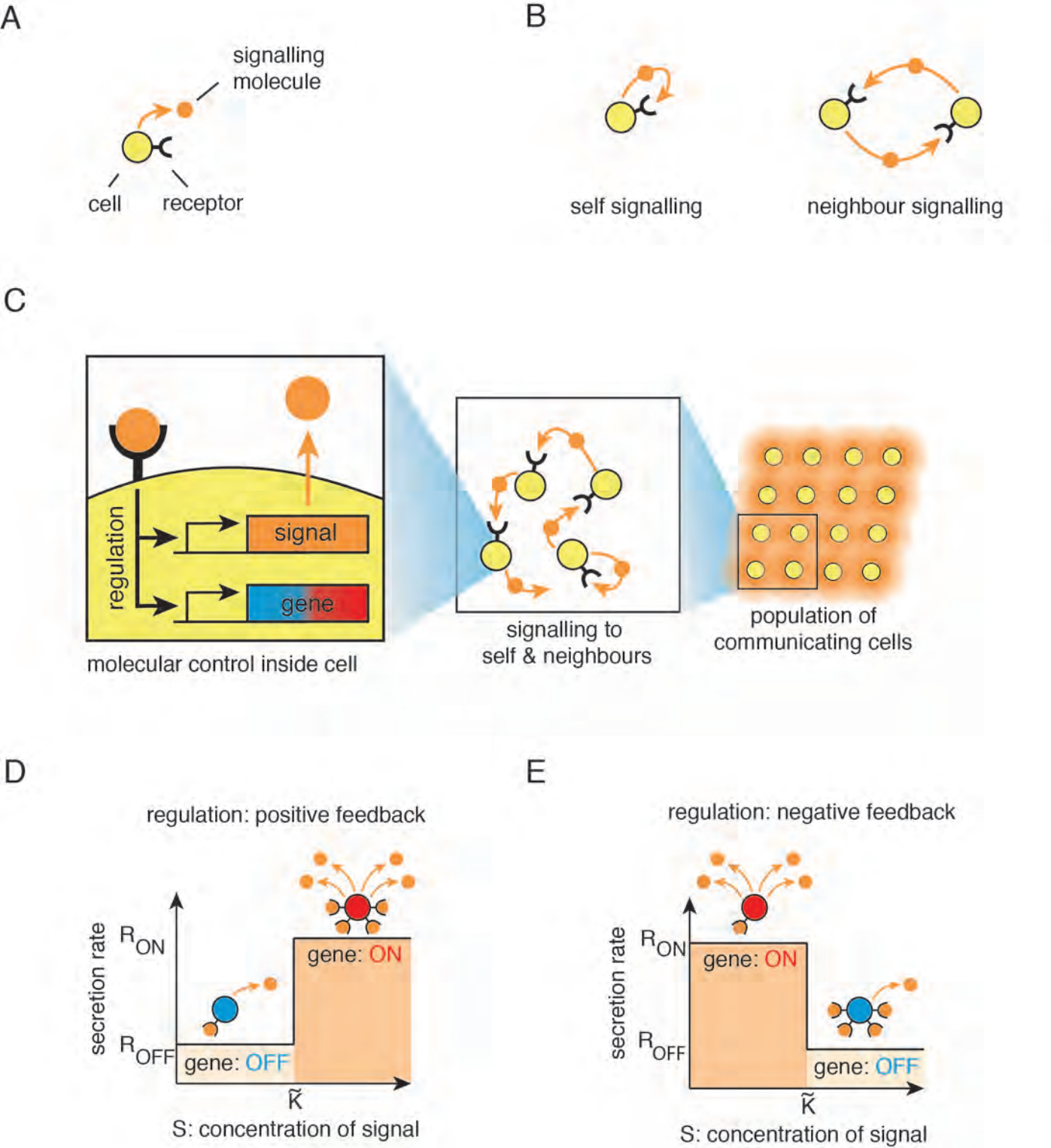

Fig. 1

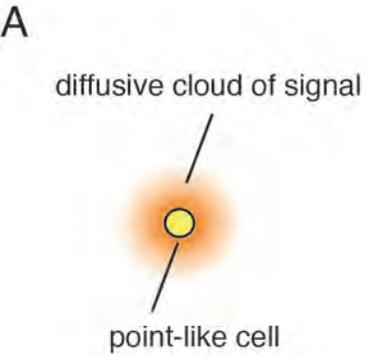
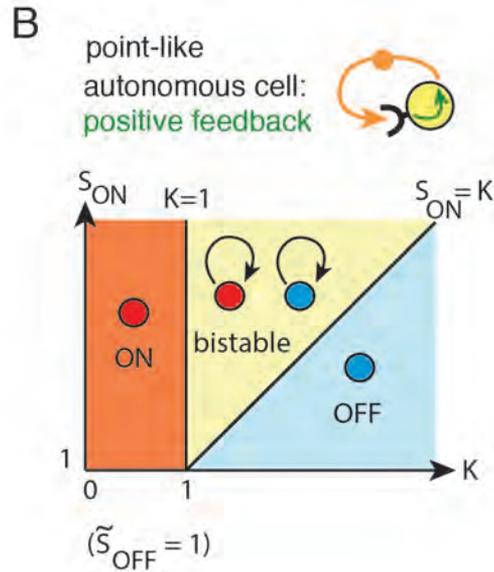
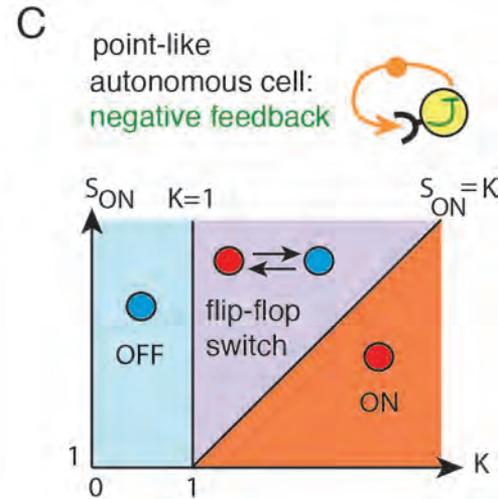
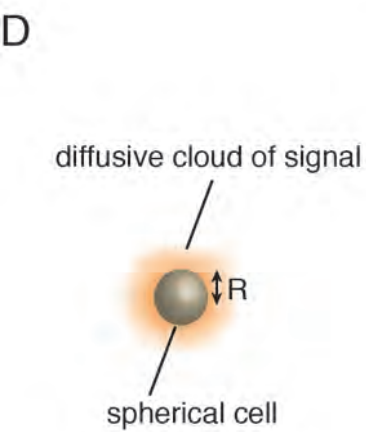
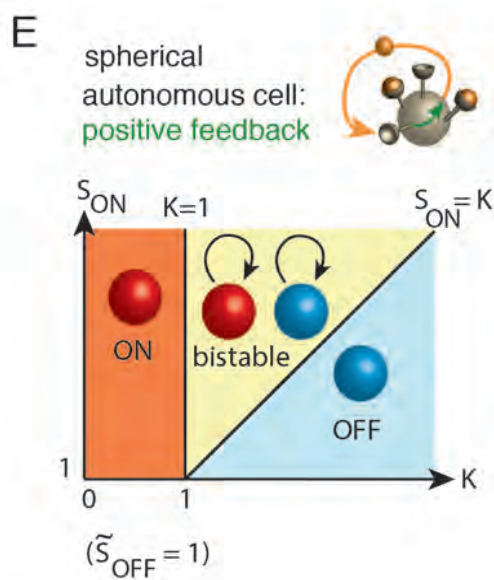
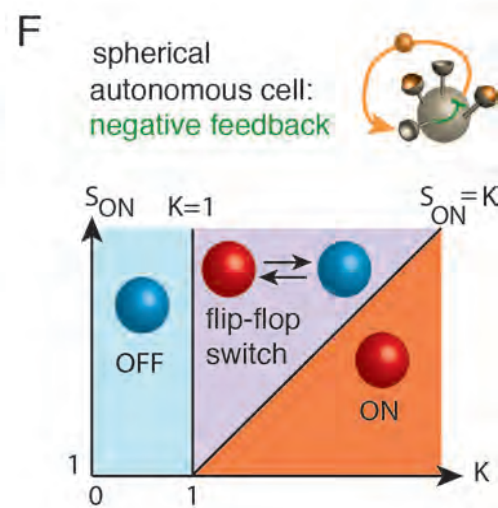

Fig. 2

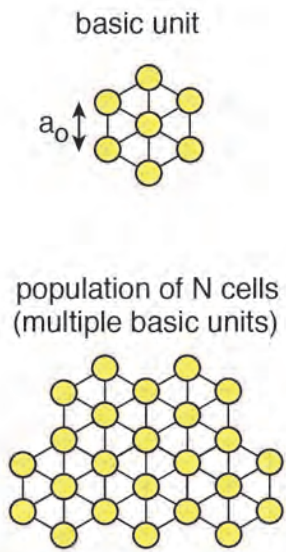
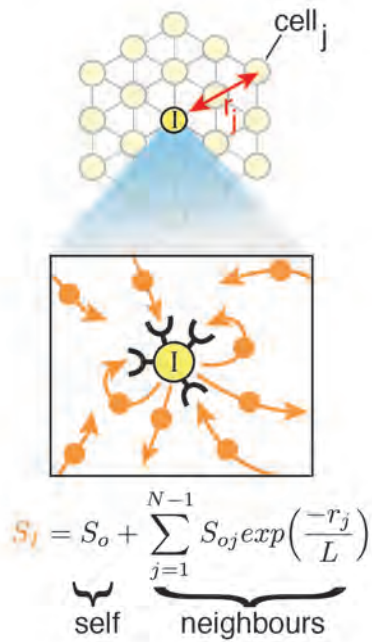
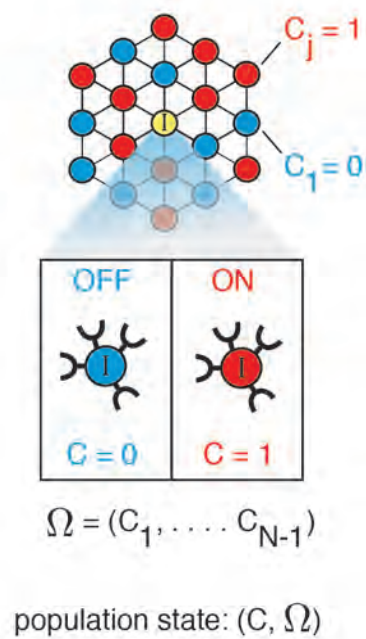
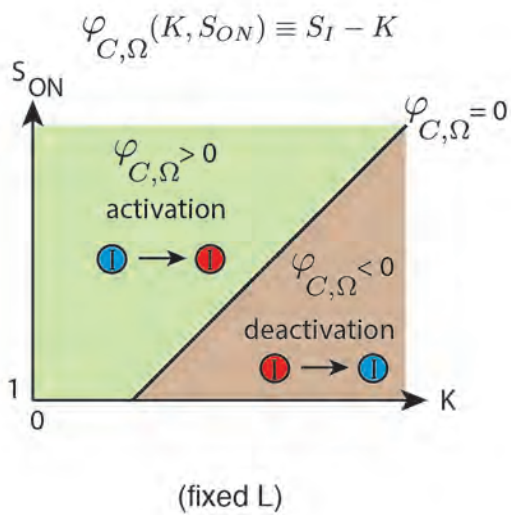
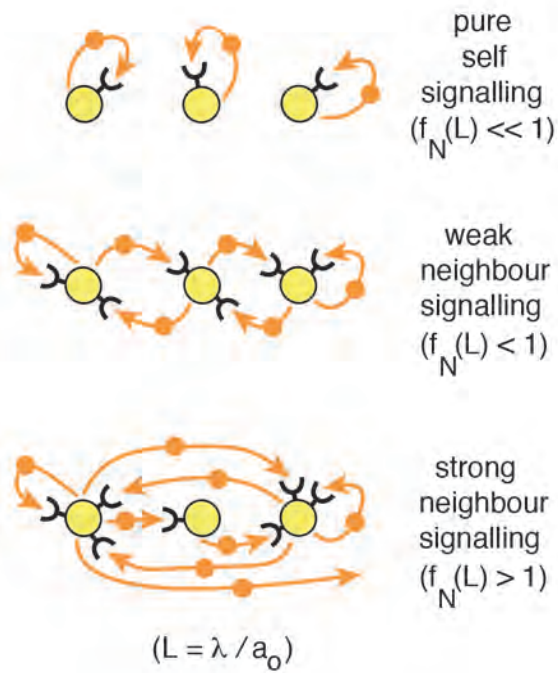

Fig. 3

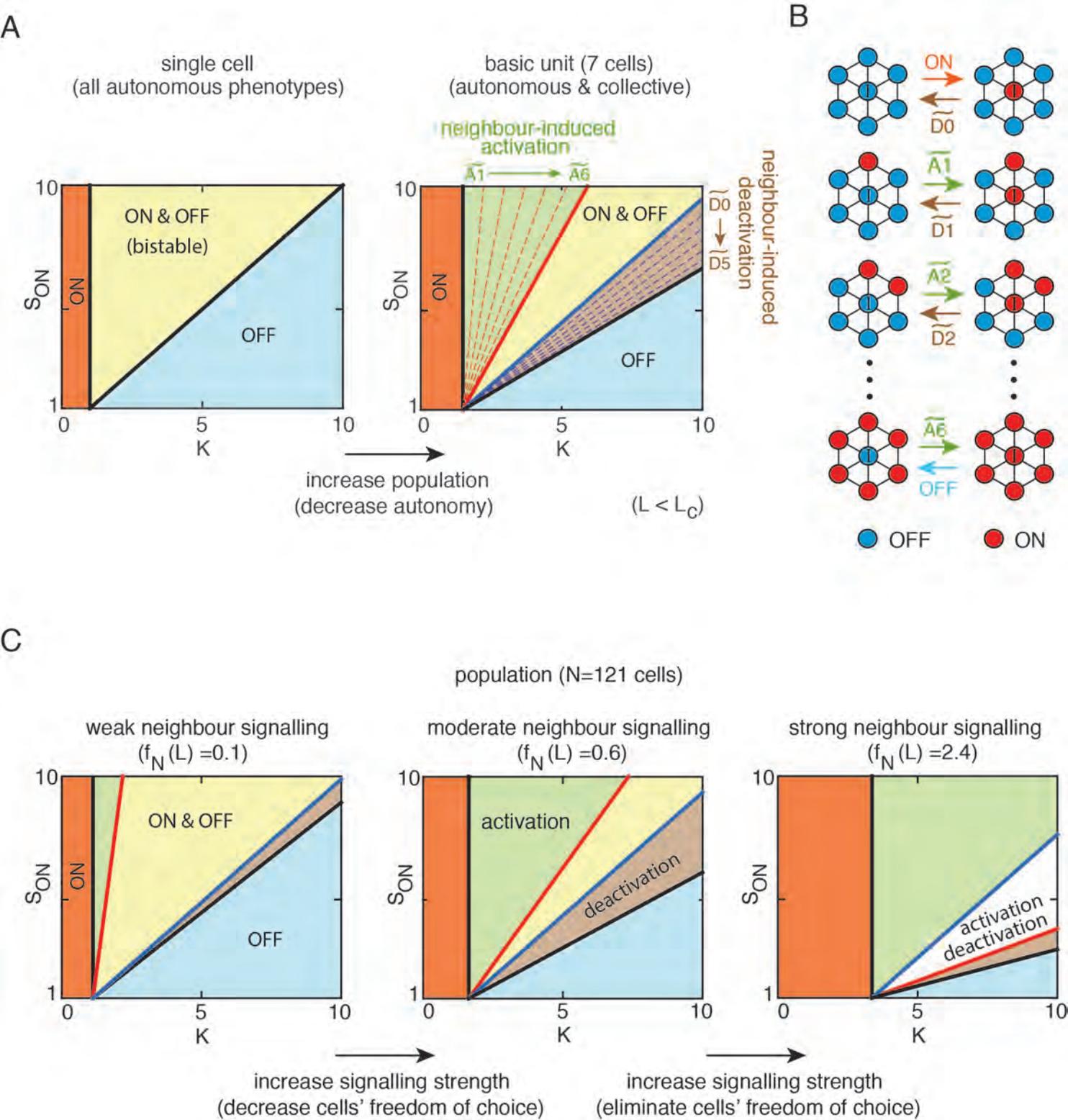

Fig. 4

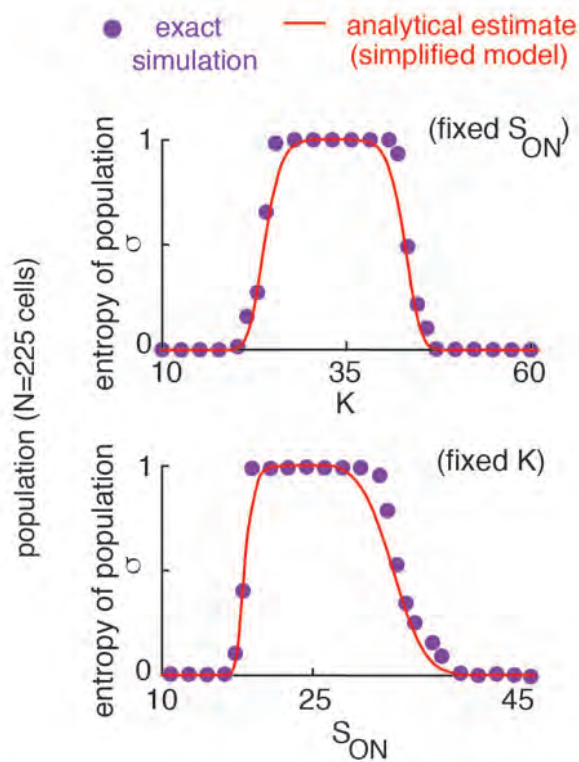
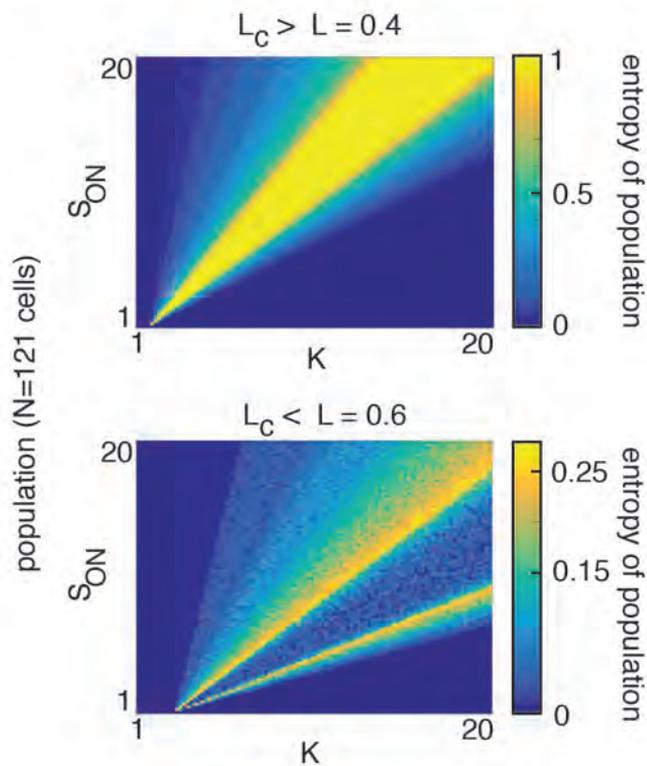
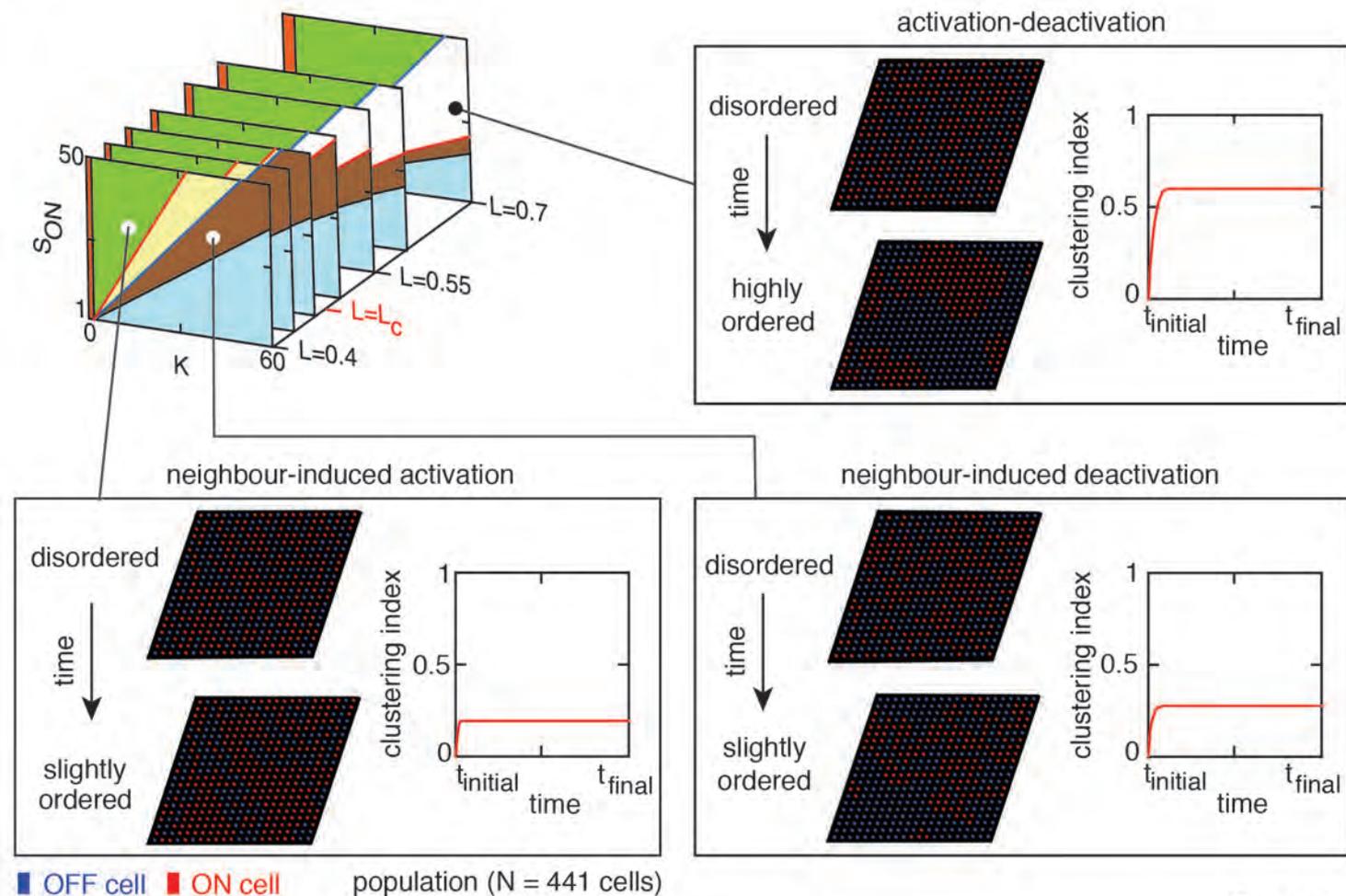

Fig. 5

## THEORETICAL METHODS

### Basic unit: Boundaries of phenotypes

The boundaries within the activation and the deactivation regions for the basic unit (Figure 4A) are given by $A_n(K,S_{ON},L)$ and $D_n(K,S_{ON},L)$ respectively:

$$A_n(K, S_{ON}, L) = S_{ON} - \frac{1}{ne^{-1/L}}K + \frac{1+(6-n)e^{-1/L}}{ne^{-1/L}}, \quad n = 1\ldots 6 \qquad [9]$$

$$D_n(K, S_{ON}, L) = S_{ON} - \frac{1}{1+ne^{-1/L}}K + \frac{(6-n)e^{-1/L}}{1+ne^{-1/L}}, \quad n = 1\ldots 6$$

with $A_0(K,S_{ON},L) = -K + 1 + 6e^{-1/L}$ and $D_0(K,S_{ON},L) = S_{ON} - K + 6e^{-1/L}$. Details are in the Supplemental Theoretical Procedures.

### Population with *N* cells: Boundaries of phenotypes

With the $A_n$ and $D_n$ defined as above, the boundaries in the phenotype diagram for *N* cells (Figure 4C) are

$$\begin{cases} A_0(K, S_{ON}, L) = 1 + f_N(L) - K \\ A_{N-1}(K, S_{ON}, L) = S_{ON} + \frac{1-K}{f_N(L)} \\ D_0(K, S_{ON}, L) = S_{ON} - K + f_N(L) \\ D_{N-1}(K, S_{ON}, L) = S_{ON} - \frac{K}{1+f_N(L)} \end{cases} \qquad [10]$$

### Definition of the clustering index

We define a clustering index $I_M$ that quantifies how closely ON cells (and thus OFF cells) are clustered together in space:

$$I_M \equiv \left[ \frac{1}{\sum_{i=1}^{N}\sum_{j=1}^{N} w_{ij}} \sum_{i=1}^{N}\sum_{j=1}^{N} w_{ij}(C_i - \overline{C})(C_j - \overline{C}) \right] \frac{N}{\sum_{i=1}^{N}(C_i - \overline{C})^2} \qquad [11]$$

Here $r_{ij}$ is the distance between *i*-th and *j*-th cells and $w_{ij} \equiv 1/r_{ij}$. $C_n$ is the state of n-th cell and $\overline{C}$ is the average of all the $C_n$'s. $I_M$ can be between 0 (spatially disordered) and 1 (spatially ordered).

## SUPPLEMENTAL INFORMATION

Supplemental Information includes Supplemental Theoretical Procedures and ten figures.

## AUTHOR CONTRIBUTIONS

T.M. and H.Y. designed the research. T.M. performed the research with guidance from H.Y. Both authors wrote the manuscript.

## ACKNOWLEDGMENTS

We thank E. Helguero, B. Doganer, A. Ravensbergen, D. Alvarez, H. Le Chenadec, and A. Raj for insightful suggestions. H.Y. is partially supported by a NWO NanoFront Grant.

**Figure captions:**

**Figure 1. From molecules to populations of cells - our bottom-up approach**

**(A)** A secrete-and-sense cell.

**(B)** Secrete-and-sense cell can signal to itself (self-signaling) and signal to its neighboring cells (neighbor-signaling).

**(C)** Outline of our bottom-up approach.

**(D)** Positive feedback regulation. If the cell senses less than the threshold concentration $\widetilde{K}$ of the signaling molecule, it is in the "OFF" state and secretes the signaling molecule at a constant rate $R_{OFF}$, otherwise the cell is "ON" and secretes the molecule at the maximal rate $R_{ON}$.

**(E)** The intracellular regulation (left panel in (*C*)) can be a negative feedback.

**Figure 2. Autonomous behaviors of an isolated cell**

**(A)** An isolated point-like secrete-and-sense cell surrounded by a diffusive cloud of the signaling molecule. The decay length $\lambda$ (equation [2]) is the radius of this diffusive cloud.

**(B)** Phenotype diagram of an isolated point-like cell with the positive feedback regulation.

**(C)** Phenotype diagram of an isolated point-like cell with the negative feedback regulation.

**(D)** An isolated spherical cell with radius $R$ surrounded by a diffusive cloud of the signaling molecule.

**(E)** Phenotype diagram of an isolated spherical cell with the positive feedback regulation.

**(F)** Phenotype diagram of an isolated spherical cell with the negative feedback regulation.

**Figure 3. Quantifying degrees of autonomy and of collectiveness**

**(A)** A "basic unit" of seven cells on a regular hexagonal lattice with an edge length $a_O$ (upper panel). Adjoining multiple basic units forms a population of *N* cells (lower panel).

**(B)** Pick any cell and call it "cell-I" (**I** for "**I**ndividual"). We focus on cell-I's loss of autonomy as we tune its communication with all the other cells. $S_I$ is the concentration of the signaling molecule on cell-I. The signaling length $L$ is the distance that the signal travels before decaying.

**(C)** Population state is denoted by a string of $2^N$ binary digits: $(C, \Omega)$, where $C$ is cell-I's state ($C=0$ if cell-I is OFF, $C=1$ if cell-I is ON) and $\Omega$ is the state of each of the N-1 neighboring cells.

**(D)** Phenotype diagram of cells with the positive feedback for a particular population state $(C, \Omega)$ and a fixed signaling length $L$.

**(E)** Tuning the "signaling strength" $f_N(L)$ (equation [7]) yields three regimes of cell-cell signaling.

**Figure 4. Populations with *N* cells and various cell-cell signaling strengths**

**(A)** Phenotype diagrams for an isolated cell with the positive feedback (left panel) and the hexagonal basic unit with a positive feedback (right panel), $L = 0.4$ ($L_c \approx 0.56$). The neighbor-induced activation region is green and the neighbor-induced deactivation region is brown. Equation [9] describes the boundary lines.

**(B)** Each region in the basic unit's phenotypic diagram (right panel, (A)) represents a state transition as shown here.

**(C)** Phenotype diagrams for a population with 121 cells (11 x 11 grid of cells) at different values of $L$, with $L_c \approx 0.47$. The neighbor-induced activation region is green and the neighbor-induced deactivation region is brown. For $L > L_c$, the "activation-deactivation" region (white region) arises.

**Figure 5. From disorder to order: Entropy of population and spatial clustering index.**

**(A)** The entropy of population (equation [8]) obtained by exact simulations (purple points) and by an analytical formula (red curve - see Supplemental Theoretical Procedures). Upper panel is for $S_{ON} = 40$ and the lower panel is for $K = 45$. Population size = 225 cells (grid of 15 x 15 cells). $L = 0.4$ ($L < L_c$).

**(B)** The entropy of population obtained by exact simulations for $L$ = 0.4 (upper panel) and $L$ = 0.6 (lower panel). Population size = 121 cells (grid of 11 x 11 cells). Sharp changes in the entropy of population occur at the boundaries between distinct phenotypic regions (compare with Figure 4C)

**(C)** Deterministic simulations of population dynamics. Population size = 441 cells (grid of 21 x 21 cells). OFF cells are blue and ON cells are red. Initial and final configurations of populations with temporal changes in the clustering index $I_M$ are shown. Results shown for activation region ($K$ = 15, $S_{ON}$=30, $L$=0.4), deactivation region ($K$=36, $S_{ON}$=30, $L$=0.4), and the activation-deactivation region ($K$ = 61, $S_{ON}$=30, $L$=0.7).

# List of Supplemental Items

**Supplemental Figures**

**Supplemental Theoretical Procedures**

**Supplemental figures**

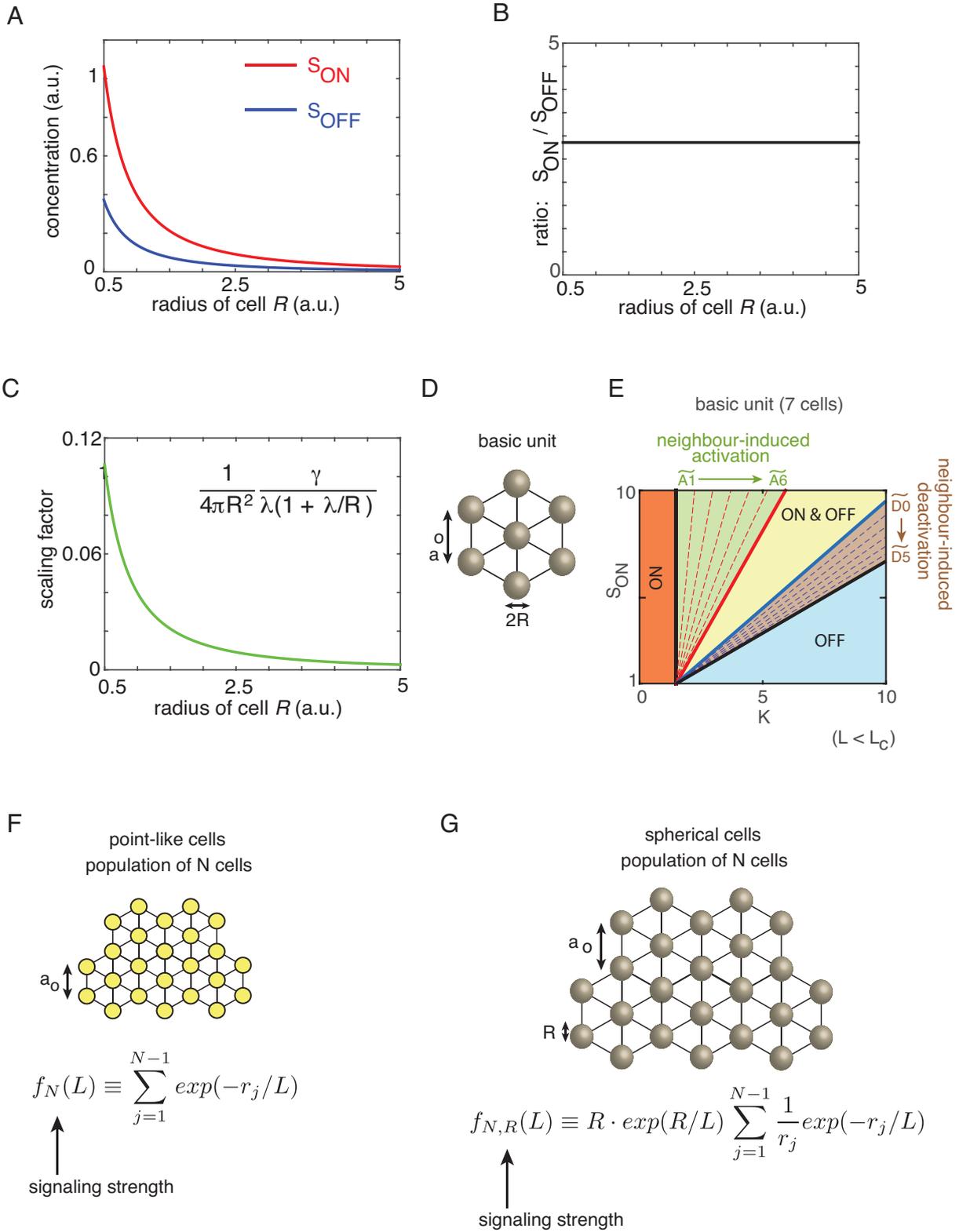

**Figure S1** *(Related to Figures 2D-2F).*

**Figure S1. Spherical cells and point-like cells have the same main features in their phenotype diagrams.**

*(Related to Figures 2D-2F)*

**(A-C) Radius of the spherical cell does not change the ratio of $S_{ON}$ and $S_{OFF}$, and thus the phenotype diagram of an isolated spherical cell is identical to the phenotype diagram of a point-like cell:** **(A)** $S_{ON}$ (red curve) and $S_{OFF}$ (blue curve) as a function of the radius $R$ of an isolated spherical cell. $S_{ON}$ and $S_{OFF}$ are the steady state concentrations on the surface of the ON cell (secreting at rate $R_{ON}$) and the OFF cell (secreting at rate $R_{OFF}$) respectively. For illustration, we plotted $S_{ON}$ and $S_{OFF}$ for $R_{ON}$ = 10, $R_{OFF}$ = 3.5, $\gamma$ = 1, and $\lambda$ =1. Increasing the cell's radius has the same effect as decreasing the secretion rates $R_{ON}$ and $R_{OFF}$ by the same amount. **(B)** The fraction $S_{ON}$ / $S_{OFF}$ remains unchanged and equals $R_{ON}$ / $R_{OFF}$ regardless of changes to the radius $R$. For this reason, we can measure all concentrations in units of $S_{OFF}$ (i.e., set $S_{OFF}$ = 1) and obtain phenotype diagrams of the isolated spherical cell that are identical to those of the isolated point-like cell (Figure 2). **(C)** $S_{ON}$ and $S_{OFF}$ are both changed by the same scaling factor.

**(D-E) The phenotype diagram of a basic hexagonal unit of spherical cells with radius $R$ is identical to that of the basic unit composed of point-like cells with minor quantitative differences.** **(D)** Basic unit composed of 3-dimensional spherical cells with radius $R$. $a_O$ is the distance between the centers of two adjacent spherical cells. **(E)** The phenotype diagram of a spherical cell with the positive feedback. This phenotype diagram is identical to the phenotype diagram of the basic unit composed of point-like cells except for two quantitative differences: (1) $S_{ON}$ for the spherical cell depends on the radius $R$, and (2) The concentration $S_I$ of the signaling molecule that cell-I (middle cell in the lattice) senses is based on a slightly different formula than that of the point-like cells (i.e., exp(-1/L) now becomes $exp(-1/2L)/2$, see supplementary text for the formula). This changes the slope of the boundaries in the phenotype diagram but the phenotypes themselves remain the same as in the point-like cells. Crucially, since we can always measure concentrations in units where $S_{OFF}$ = 1 as we did in the case of point-like cells, the phenotype diagram (B) is invariant under changes in the value of the radius $R$ as long as we reset the unit of concentration so that $S_{OFF}$ = 1 whenever we change the value of $R$.

**(F-G) The phenotype diagrams of a population of $N$ spherical cells with radius $R$ are identical to those of the population of $N$ point-like cells with minor quantitative differences:** **(F)** The signaling strength function $f_N(L)$ for a population of $N$ point-like cells (equation [7] in the main text). This function completely determines the boundaries of the phenotype diagrams as a function of the signaling length $L$. **(G)** The signaling strength function $f_{N,R}(L)$ for a population of $N$ spherical cells with radius $R$. Note that the two signaling strength functions, $f_{N,R}(L)$ and $f_N(L)$, are similar to each other. The two main differences are:

(1) The $1/r_j$ factor in front of the $\exp(-r_j / L)$ means that the contribution of each spherical cell to the overall signaling strength is the contribution of each point-like cell modulated by $1/r_j$, and (2) the radius of the cells modulates the combined contributions from every cell. Despite these two quantitative differences, the procedure for computing the phenotype diagrams for $N$ spherical cells is the same as the procedure for computing the phenotype diagrams of $N$ point-like cells. Importantly, the phenotype diagrams for $N$ spherical cells have the same phenotypes as the $N$ point-like cells. There are still three regimes: (1) $f_{N,R}(L) \ll 1$ (pure self signaling), (2) $f_{N,R}(L) < 1$ (weak neighbor signaling), and (3) $f_{N,R}(L) > 1$ (strong neighbor signaling). But now the value of the critical length $L_c$ depends on both $R$ and $N$. Aside from these quantitative differences, the phenotype diagrams for $N$ spherical cells in each of these three regimes are exactly identical to the phenotype diagrams of the $N$ point-like cells (Figure 4C).

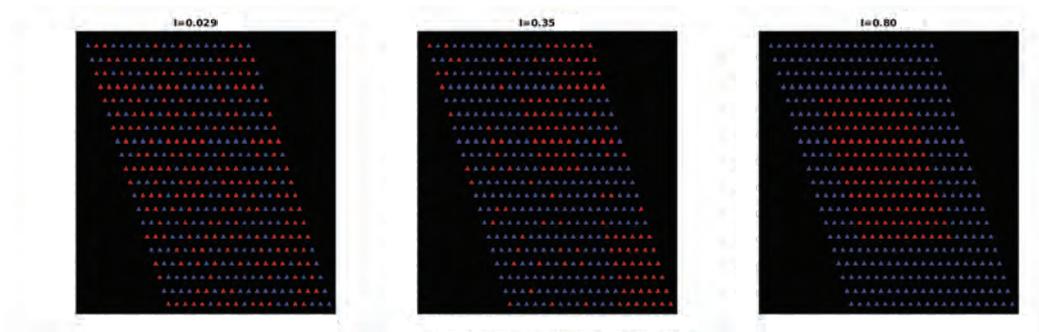

**Figure S2. Examples of different clustering index values.**

*(Related to Figure 5)*

Disordered population ($I_M \sim 0.029$, left panel), a population with a marginal spatial ordering ($I_M \sim 0.35$, middle panel), and a population with a high spatial order ($I_M \sim 0.80$, right panel).

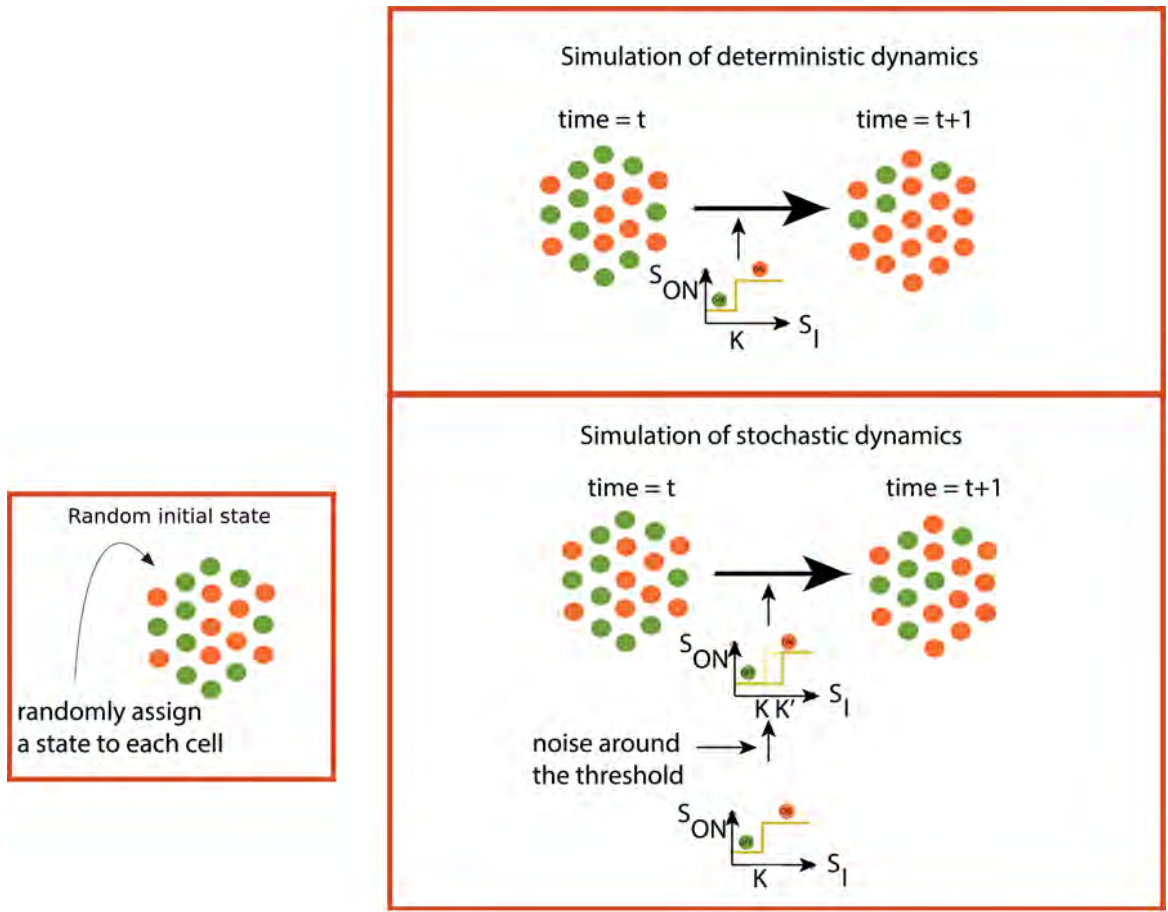

**Figure S3. Schematic of deterministic and stochastic simulations of population dynamics.**

*(Related to Figure 5)*

In both deterministic and stochastic simulations, cells are placed on a lattice. Schematic is shown only for cells with the positive feedback since the negative feedback case follows the same principle. Each cell's state is randomly chosen to be either ON (orange) or OFF (green). In the deterministic simulation, the current state of population (C, Ω) completely determines its next state (determined by the $S_I$, equation [6], applied to each cell). In the stochastic simulation, we still use equation [6] to compute $S_I$ but each cell can make an error in sensing the concentration. This occurs if the $S_I$ at the cell of interest is near the threshold concentration $K$. We define a range of concentration around $K$, (K - δK, K+ δK). If the $S_I$ is in this interval, an OFF cell can turn ON even if the true $S_I$ below the threshold. Similarly, an ON cell can turn OFF if $S_I$ is in the interval (K - δK, K+ δK). This is similar to making the step-function that represents the positive feedback smoother (i.e., lowering the Hill coefficient to a finite value so that a sigmoidal curve replaces the step function). In both simulations, time is measured in discrete steps. Each time step represents a change in the population state. We run the simulations until the population reaches an equilibrium state.

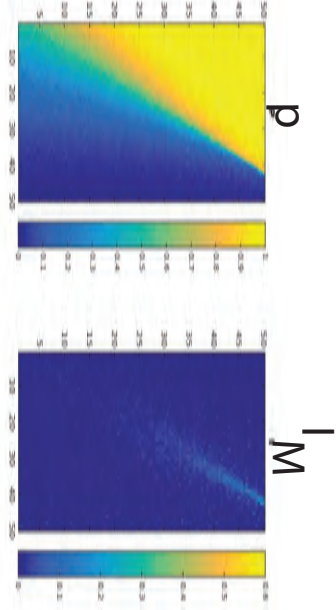
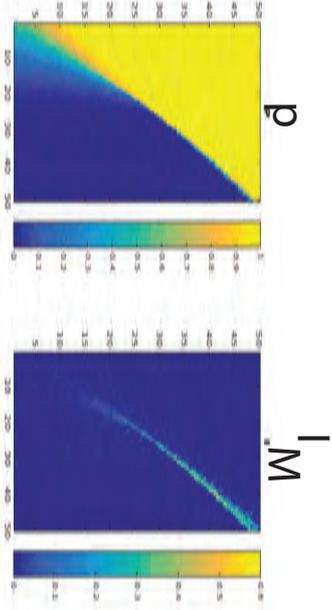
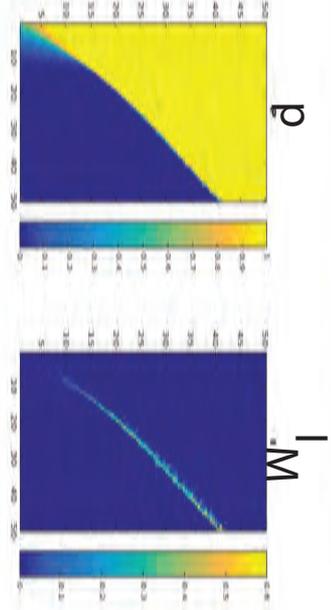
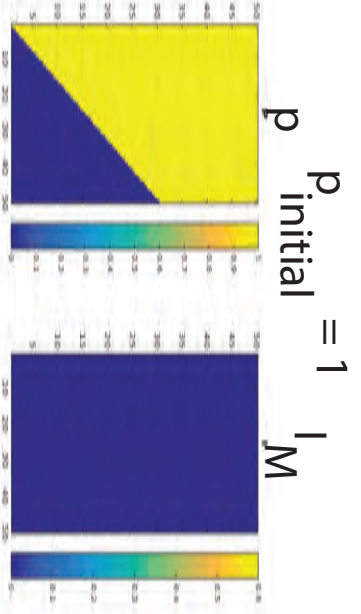
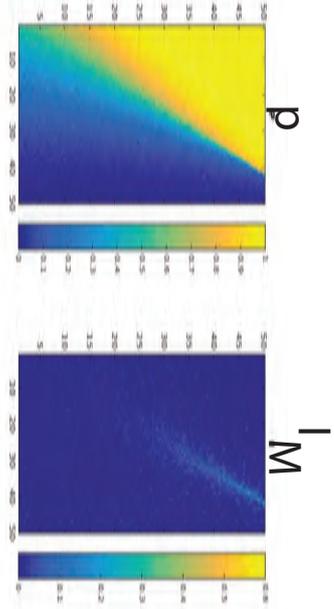
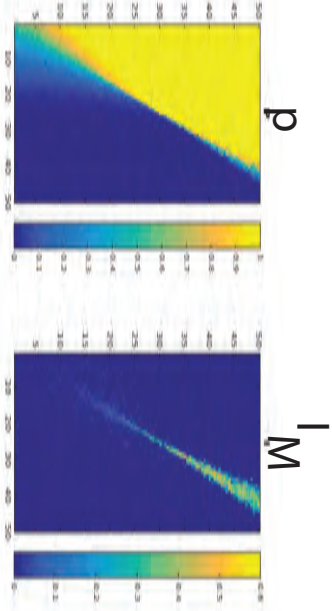
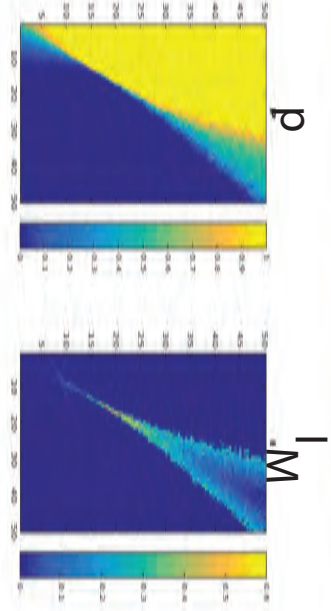
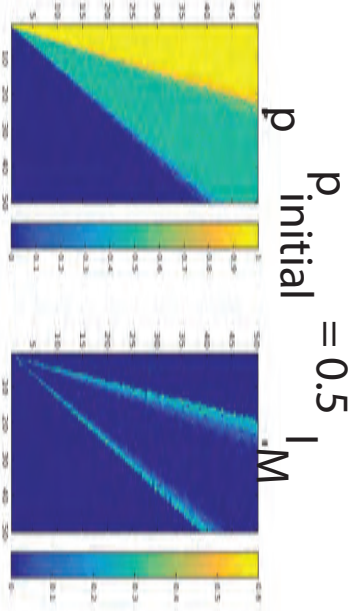
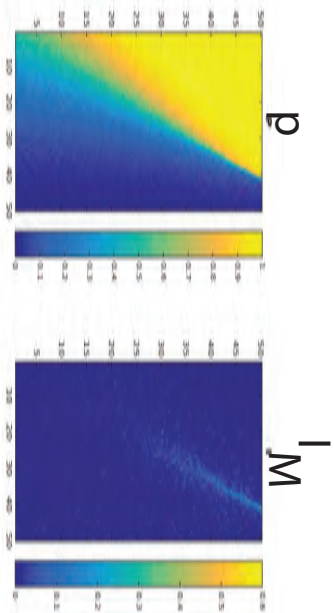
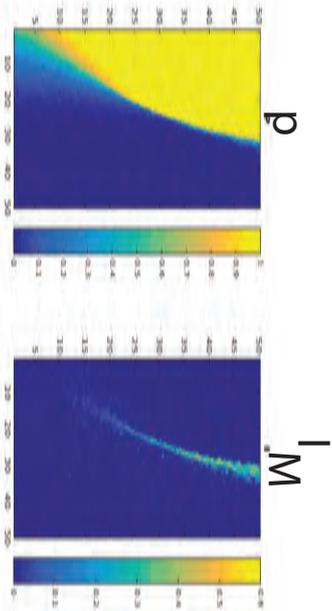
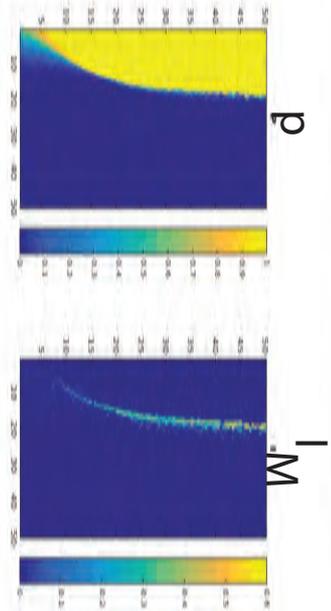
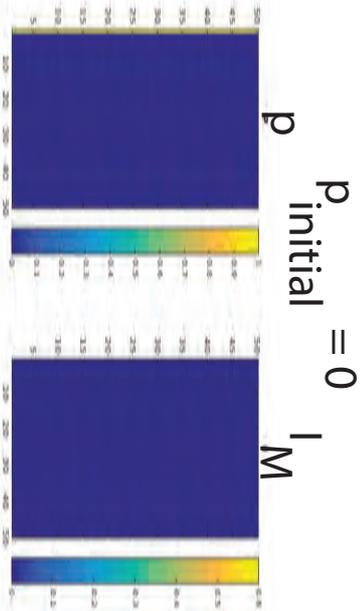

**Figure S4. Deterministic and stochastic simulations ($L$ = 0.4, $L < L_c$).**

*(Related to Figure 5)*

Left column represents deterministic (top row) and stochastic ($2^{nd} \sim 4^{th}$ rows) simulations (see Figure S3) that started with initial fraction $p_{initial}$ of ON cells equal to 0 (i.e., all cells are OFF). Middle column represents simulations that started with initial fraction $p_{initial}$ of ON cells equal to 0.5 (i.e., half the cells are ON). Last column represents simulations that started with initial fraction $p_{initial}$ of ON cells equal to 1 (i.e., all the cells are ON). All the heatmaps represent the final, equilibrium state. In all the heat maps, the horizontal axis denotes values of $K$ ($0 < K < 50$) and the vertical axis denotes values of $S_{ON}$ ($0 < S_{ON} < 50$). $p$ is the fraction of ON cells in the population. $I_M$ is the clustering index (defined in equation [13] in the main text). In all the heat maps of $p$, darkest blue represents $p = 0$ (all cells are OFF) and darkest yellow represents $p = 1$ (all cells are ON). In all the heat maps of $I_M$, darkest blue represents $I_M = 0$ (random, spatially disordered population) and darkest yellow represents $I_M = 0.6$ (more spatially ordered population).

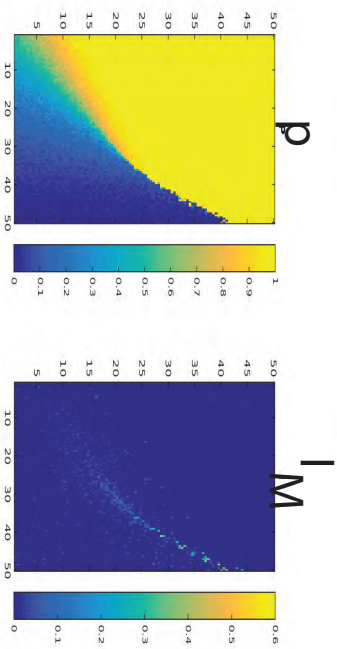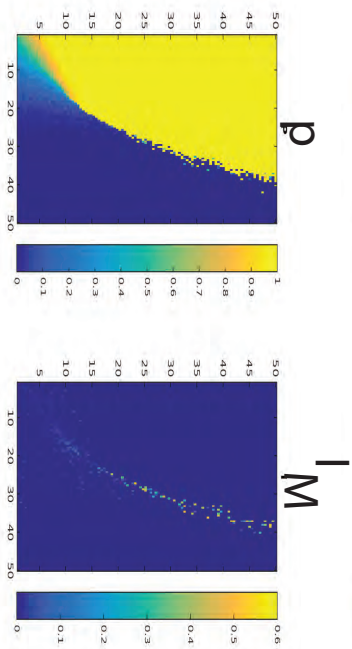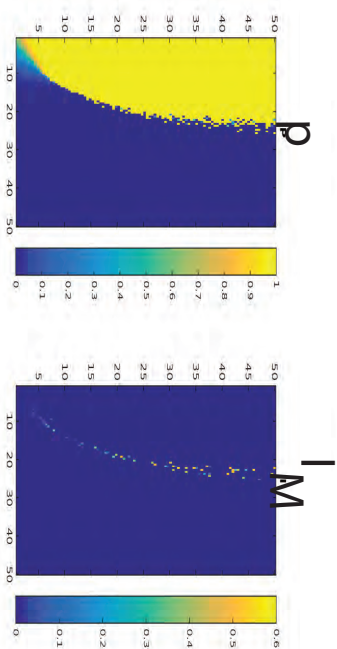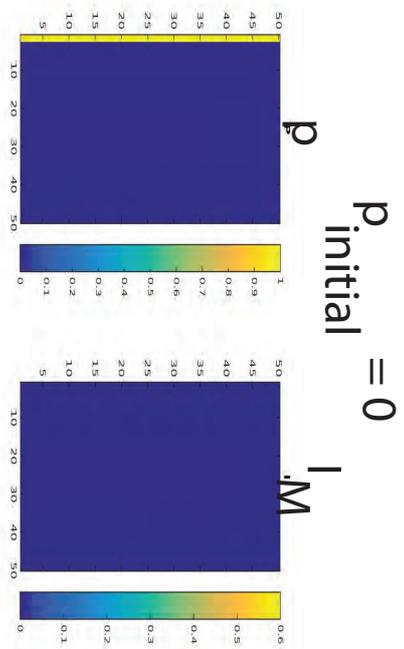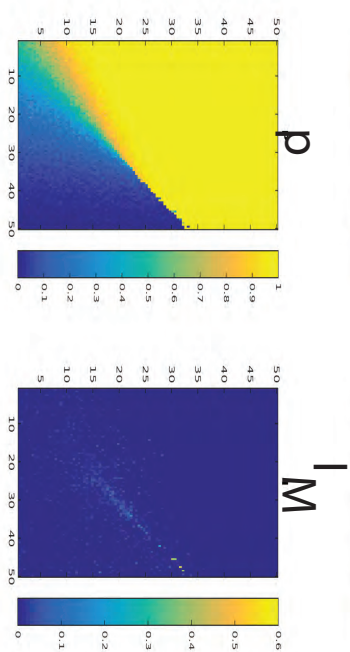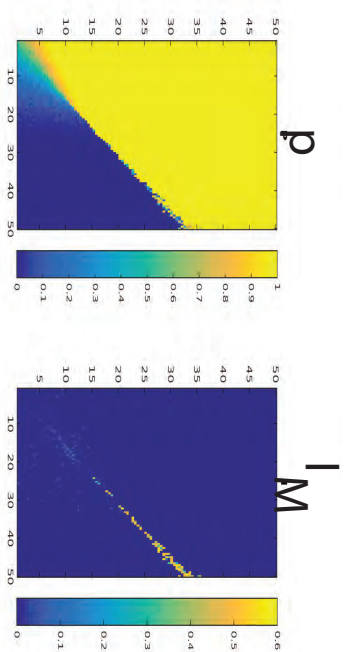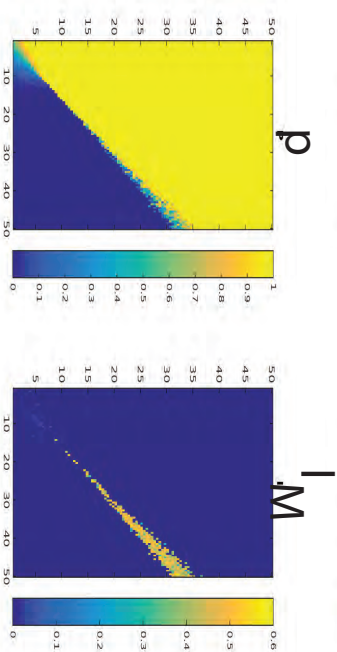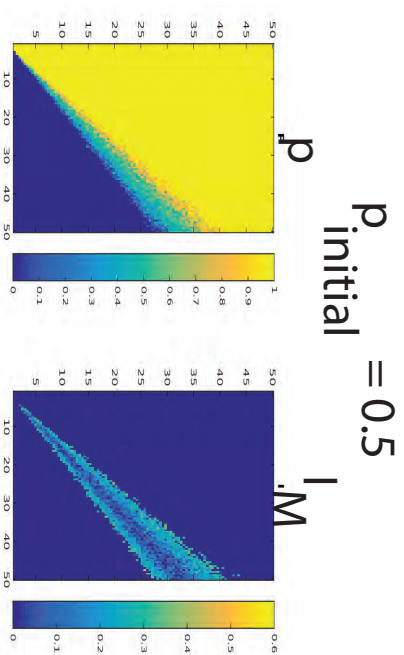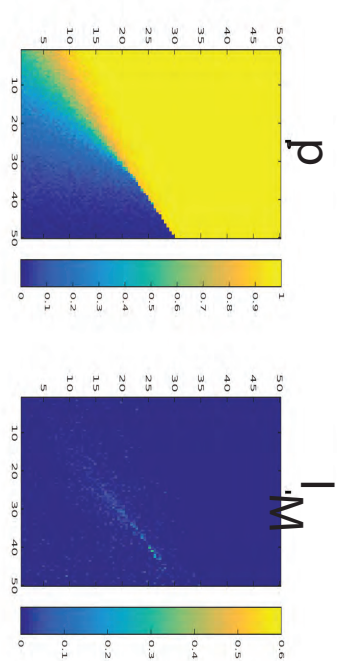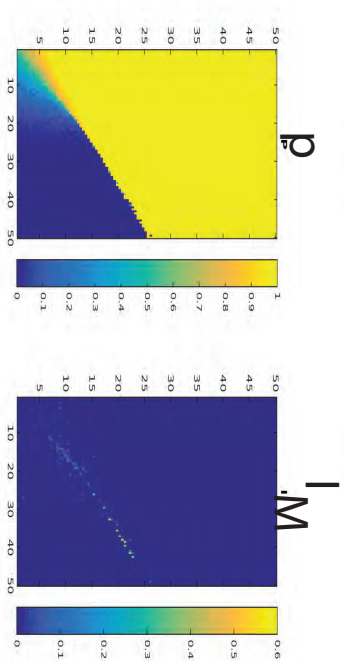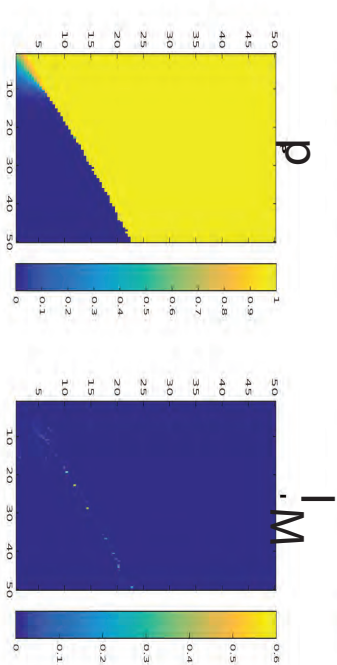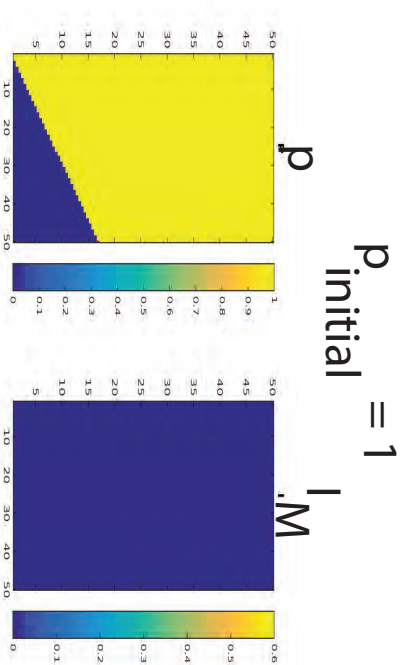

**Figure S5. Deterministic and stochastic simulations ($L = 0.6$, $L > L_c$).**

*(Related to Figure 5)*

Left column represents deterministic (top row) and stochastic (2nd ~ 4th rows) simulations (see Figure S3) that started with initial fraction $p_{initial}$ of ON cells equal to 0 (i.e., all cells are OFF). Middle column represents simulations that started with initial fraction $p_{initial}$ of ON cells equal to 0.5 (i.e., half the cells are ON). Last column represents simulations that started with initial fraction $p_{initial}$ of ON cells equal to 1 (i.e., all the cells are ON). All the heat maps represent the final, equilibrium state. In all the heat maps, the horizontal axis denotes values of $K$ ($0 < K < 50$) and the vertical axis denotes values of $S_{ON}$ ($0 < S_{ON} < 50$). $p$ is the fraction of ON cells in the population. $I_M$ is the clustering index (defined in equation [13] in the main text). In all the heat maps of $p$, darkest blue represents $p = 0$ (all cells are OFF) and darkest yellow represents $p = 1$ (all cells are ON). In all the heat maps of $I_M$, darkest blue represents $I_M = 0$ (random, spatially disordered population) and darkest yellow represents $I_M = 0.6$ (more spatially ordered population).

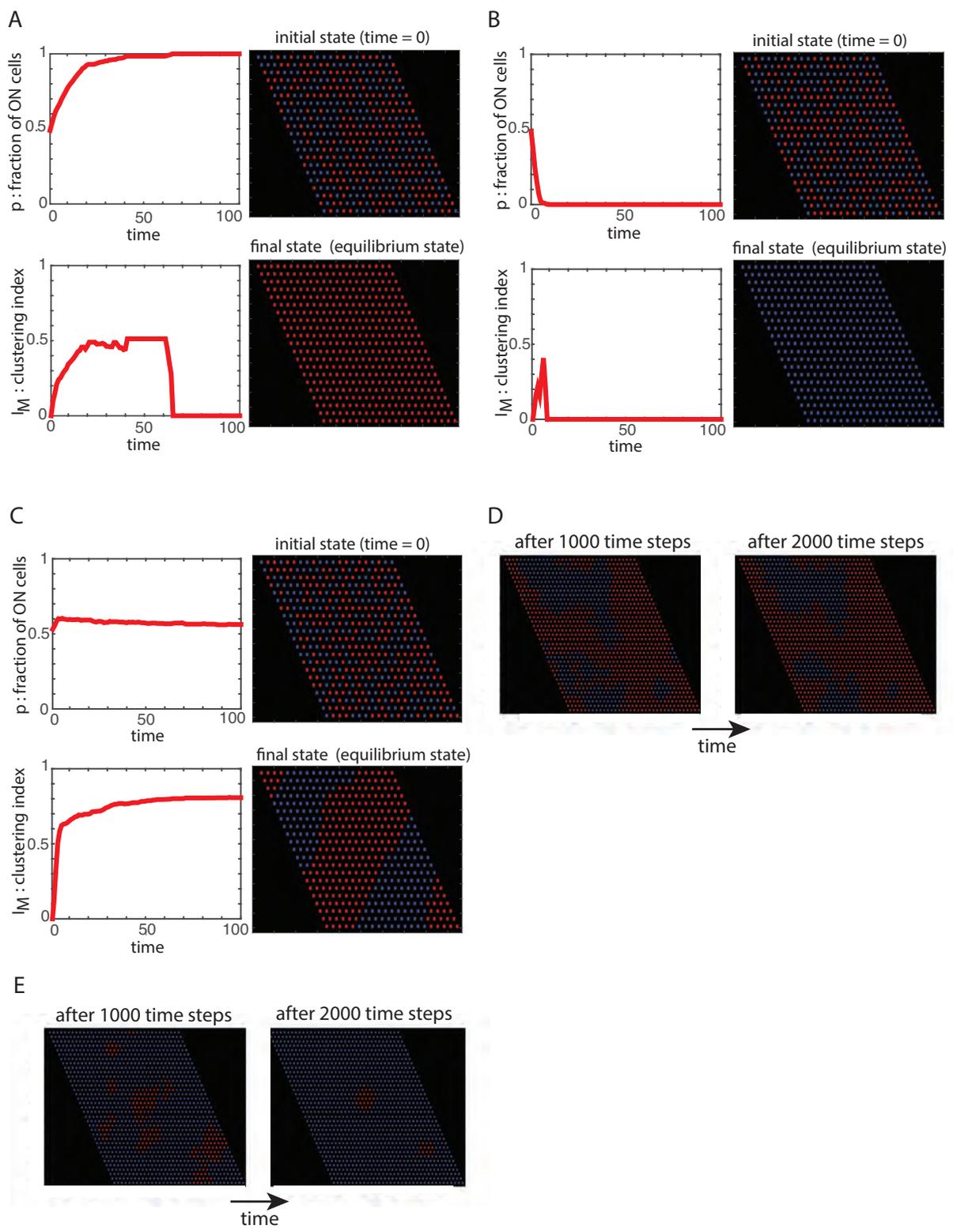

**Figure S6.** *(Related to Figure 5)*

**Figure S6. Population dynamics**

*(Related to Figure 5)*

**(A) Noise driven activation ($L < L_c$):** Typical dynamics. Stochastic simulation (see Fig. S3) performed for a population of 441 cells (grid of 21 x 21 cells) that have positive feedback and are in the "activation region" of the phenotype diagram ($K$ = 15, $S_{ON}$ = 30, $L$ = 0.4, $L < L_c$). The population starts with the fraction $p$ = 0.5 ON cells without any spatial order $I_M \sim 0$. Then it reaches an equilibrium state ("final state") over time according to the stochastic dynamics previously described. Time is measured in discrete steps. Each time step represents a change in the population state. The final state has every cell ON (thus $p$ = 1) and a trivial spatial order (i.e., no islands, stripes, or patterns). Thus $I_M$ = 0 in the final state. Note that noise drives the population into the extreme ON state (i.e., everyone is ON) whereas in the deterministic simulation, $p$ does not reach 1 for many initial population states.

**(B) Noise driven deactivation ($L < L_c$):** Typical dynamics. Stochastic simulation (see Fig. S3) performed for a population of 441 cells (grid of 21 x 21 cells) that have positive feedback and are in the "deactivation region" of the phenotype diagram ($K$ = 36, $S_{ON}$ = 30, $L$ = 0.4, $L < L_c$). The population starts with the fraction $p$ = 0.5 ON cells without any spatial order $I_M \sim 0$. Then it reaches an equilibrium state ("final state") over time according to the stochastic dynamics previously described. Time is measured in discrete steps. Each time step represents a change in the population state. The final state has every cell OFF (thus $p$ = 0) and a trivial spatial order (i.e., no islands, stripes, or patterns). Thus $I_M$ = 0 in the final state. Note that noise drives the population into the extreme OFF state (i.e., everyone is OFF) whereas in the deterministic simulation, $p$ does not reach 0 for many initial population states.

**(C) Noise driven simultaneous activation and deactivation ($L > L_c$):** Typical dynamics. Stochastic simulation (see Fig. S3) performed for a population of 441 cells (grid of 21 x 21 cells) that have positive feedback and are in the "activation & deactivation region" of the phenotype diagram ($K$ = 61, $S_{ON}$ = 30, $L$ = 0.7, $L > L_c$). The population starts with the fraction $p \sim 0.5$ ON cells without any spatial order $I_M \sim 0$. Then it reaches an equilibrium state ("final state") over time according to the stochastic dynamics previously described. Time is measured in discrete steps. Each time step represents a change in the population state. The final state has nearly half of the cells ON and the other half of the cells in the OFF state (thus $p \sim 0.55$). The final population is highly ordered in space ($I_M \sim 0.8$). Note that noise can drive the population into a state that has a higher spatial order (i.e., larger $I_M$) than the spatial order that the population can typically achieve when it evolves deterministically.

**(D-E) Initially disordered population can evolve over time to form highly ordered patterns that are stable over a long time:** Pictures from stochastic simulations (see Fig. S3). All populations have 1225 cells (grid of 35 x 35 cells). All populations initially start with

random ($I_M \sim 0$) spatial arrangement of ON and OFF cells, and with $p$=0.5 (i.e. 50% of the cells are initially ON, the other 50% of the cells are initially OFF). **(D)** Population state after 1000 time steps (left panel) and after 2000 time steps (right panel). $L$ = 0.6, $K$ = 45, $S_{ON}$ = 30. **(E)** Population state after 1000 time steps (left panel) and after 2000 time steps (right panel). $L$ = 0.6, $K$ = 47.5, $S_{ON}$ = 30.

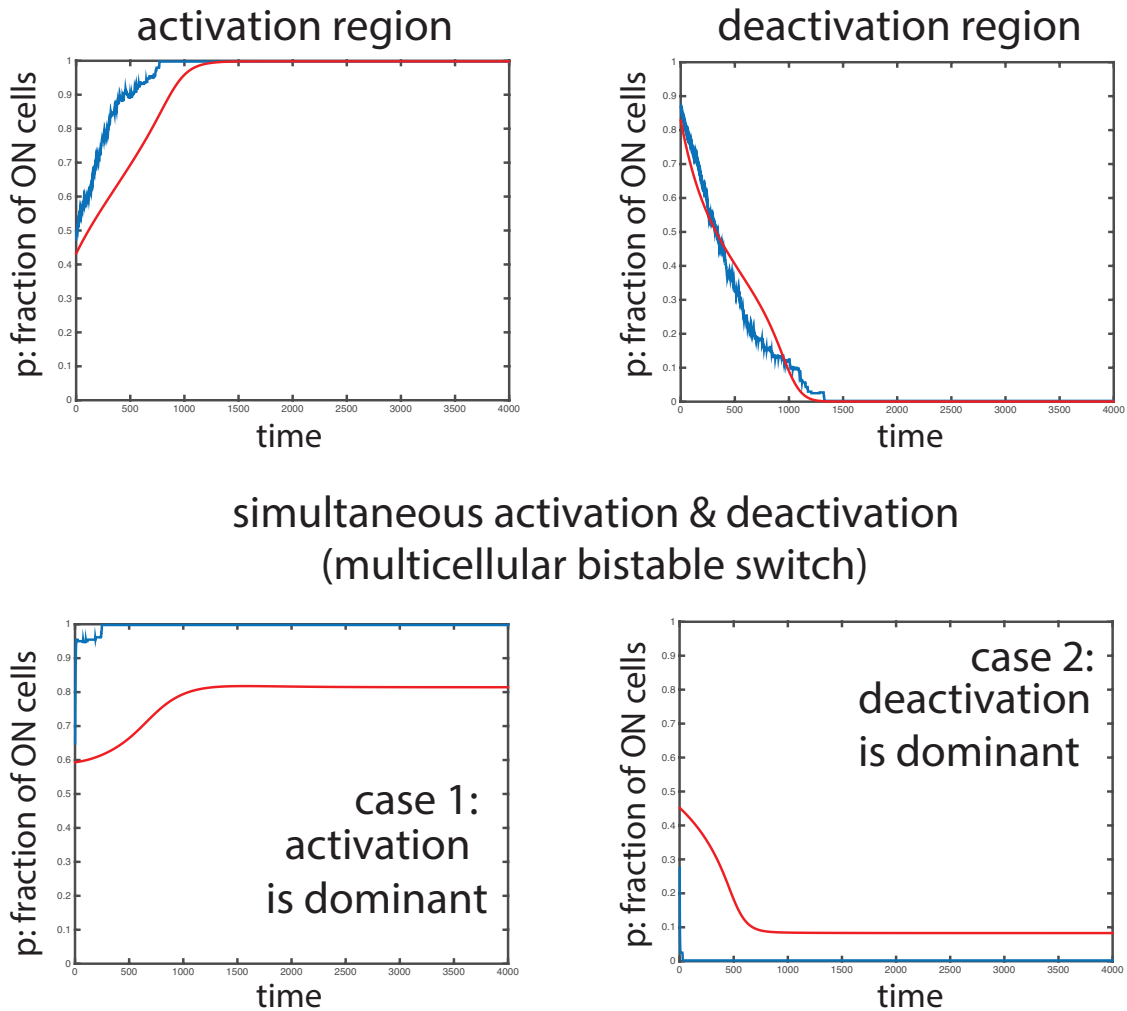

**Figure S7. Comparison between the population dynamics dictated by the mean-field model and the deterministically simulated population dynamics.**
*(Related to Figure 5)*

Red curve is the fraction *p* of ON cells over time dictated by the discrete version of the mean-field model (section 3a of Supplementary Information). Blue curve is the exact value of *p* obtained through the deterministic simulations (Figure S3). The mean-field model recapitulates the main qualitative features of the temporal changes in *p* seen in the simulations. We observed larger deviations between the two versions of the *p*'s when the initial spatial order (i.e., the initial clustering index $I_M$) was closer to 1. This deviation quantifies the effect of spatial ordering of cell states on the dynamics of the whole population. Note that in the case of cells in the region of simultaneous activation and deactivation, we see that either the entire population tends towards everyone turning ON or towards everyone turning OFF. This depends on the initial *p* and the initial spatial arrangement of ON cells. The two cases (case 1 & case 2) shown here illustrates why we can call this region a region of "multicellular bistability".

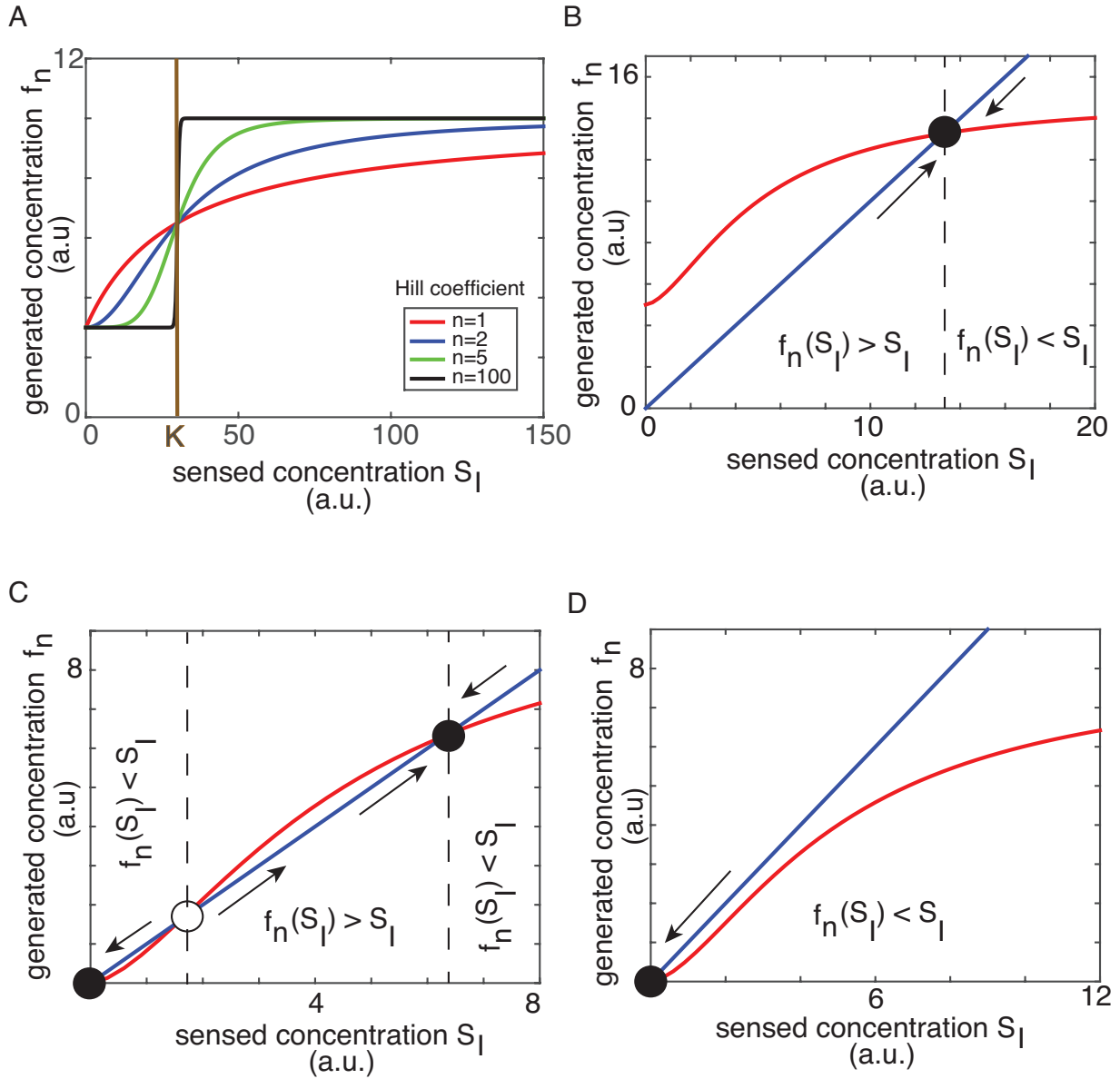

**Figure S8.**

*(Related to Figure 1)*

**(A) Sigmoidal functions with different Hill coefficients that describe the secreted concentration:** The cell secretes signaling molecules at a rate that is a sigmoidal function of the concentration $S_I$ that it is sensing at a given moment. Since the concentration created on the cell surface is directly proportional to the secretion rate, the created concentration on the cell surface $f_n$ after some fixed time interval $\delta t$ is also a sigmoidal function of the concentration $S_I$ that the cell is initially sensing. Here we have plotted four different $f_n$'s. They are for *n*=1, 2, 5, and 100. For all of them, we used *K*=30, $S_{OFF}$ = 3, and $S_{ON}$ = 10.

**(B-D) Isolated cell with a positive feedback and a finite Hill coefficient:** Red curves are the sigmoidal functions $f_n(S_I)$ with a Hill coefficient *n* that describe the secreted concentration of the signaling molecule (proportional to the sigmoidal function that describes the secretion rate) as a function of the concentration $S_I$ that the cell senses (see Supplementary text). Blue (diagonal) line represents the sensed concentration $S_I$. Note that an isolated cell cannot sense more than what it can generate through its own secretion. This is the logic behind the arrows. The arrows represent temporal evolution of the cell's state. Black circles represent stable equilibrium states. The open circle represents an unstable equilibrium state. **(B)** Self activation phenotype occurs (used $S_{OFF}$ = 5, $S_{ON}$ = 15, $K$ = 5, $n$ = 1.6). This is analogous to the turning "ON" phenotype of the cell with a positive feedback and an infinite Hill coefficient. **(C)** Bistability occurs (used $S_{OFF}$ = 0, $S_{ON}$ = 10, $K$ = 4.5, $n$ = 1.6). This is analogous to the "bistability" phenotype of the cell with a positive feedback and an infinite Hill coefficient. **(D)** Self deactivation phenotype occurs (used $S_{OFF}$ = 0, $S_{ON}$ = 8, $K$ = 5, $n$ = 1.6). This is analogous to the turning "OFF" phenotype of the cell with a positive feedback and an infinite Hill coefficient.

**Figure S9. Phenotype diagrams for a population of *N* cells with a finite Hill coefficient (Weak neighbor communication: *L* < *L_C*; signaling strength $f_N(L)$ = 0.5).**

*(Related to Figures 3D and 3E)*

We numerically computed the phenotype diagrams for a population *N* cells with a positive feedback and a Hill coefficient *n* by devising a computational algorithm (see supplementary text). We used four different values for the Hill coefficients: (1) a "low" value (*n*=1), (2) an "intermediate" value (*n*=1.5), (3) a "high" value (*n*=2), and (4) a "near infinite" value (*n*=40). To obtain some intuition for our numerical work, we obtained phenotype diagrams for four initial population states: (1) cell-I is initially OFF and everyone else is initially ON, (2) cell-I is initially OFF and everyone else is initially OFF, (3) cell-I is initially ON and everyone else is initially ON, and (4) cell-I is initially ON and everyone else is initially OFF. To obtain each phenotype diagram, we fixed the Hill coefficient *n* to be one of the four values mentioned above. Then we simulated our computational algorithm for each value of (*K*, $S_{ON}$), which used an iterative algorithm that we designed (we used 50 iterations for each value of (*K*, $S_{ON}$)). Then we used a color to represent a "phenotype score" (see Supplementary text). The phenotype score is necessary because the finite Hill coefficient allows for cells to be in a continuum of states between OFF and ON (instead of the binary ON and OFF states). Doing this for a wide range of values for *K* and $S_{ON}$ resulted in a heat map (phenotype diagram). We used a color spectrum that starts from a pure red (representing a phenotype score of 1) and ends in a pure blue (representing a phenotype score of 0). A pure red represents cell-I being in the ON-state while a pure blue represents cell-I being in the OFF-state. An intermediate color such as green represents the case in which the cell-I is in between the ON and OFF (i.e., partially ON). A color that is closer to red means that cell-I is closer to being ON while a color that is closer to blue means that cell-I is closer to being OFF. Note that the "near infinite" Hill coefficient is nearly identical to the phenotype diagram for cell-I (Figure 3D). Moreover, except for the appearance of intermediate states (e.g., green regions), we find that the main qualitative features of the phenotype diagrams for cells with a finite Hill coefficient are essentially identical to the main features of the phenotype diagrams for cells with an infinite Hill coefficient (Figure 3D).

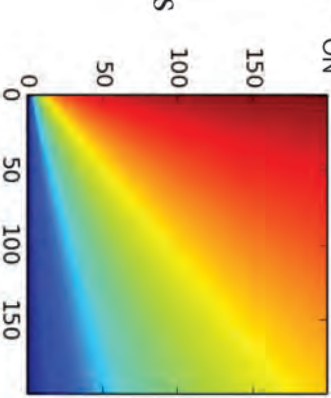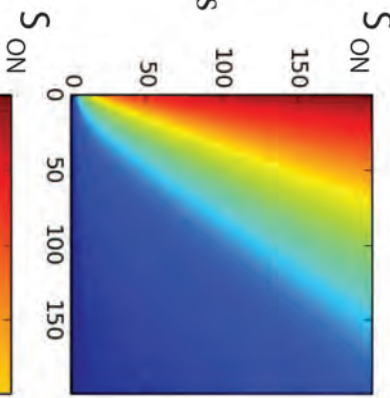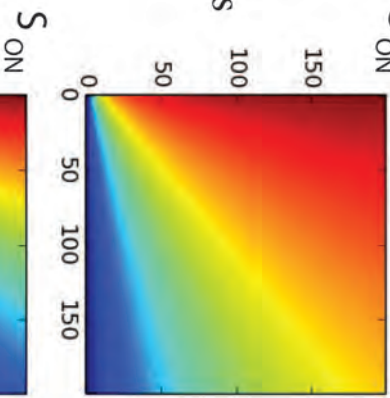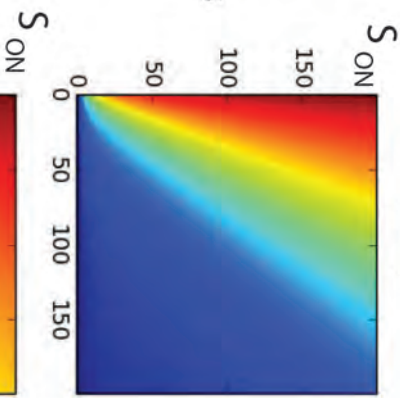
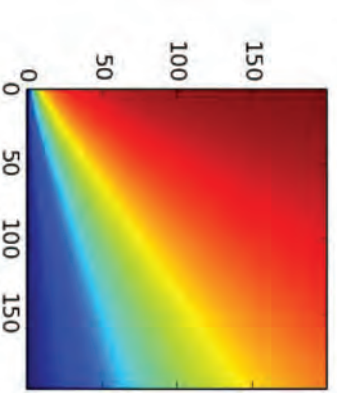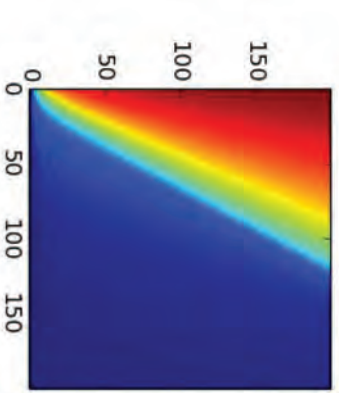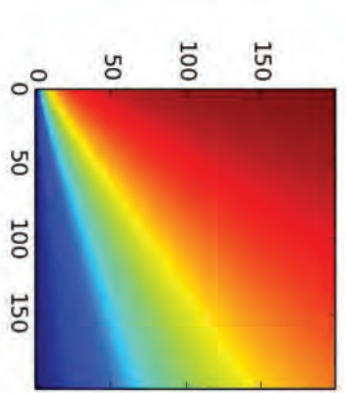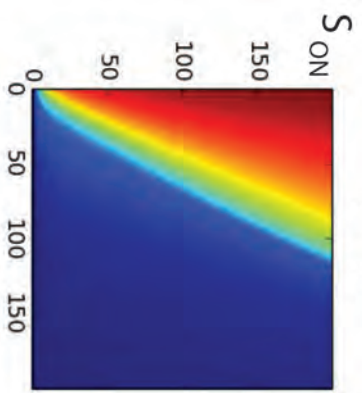
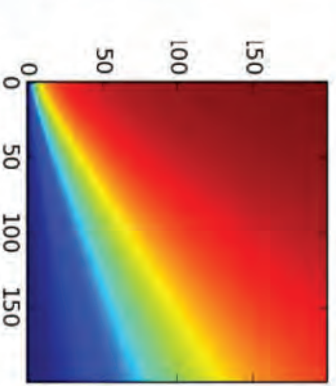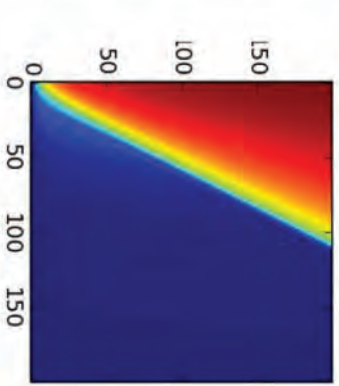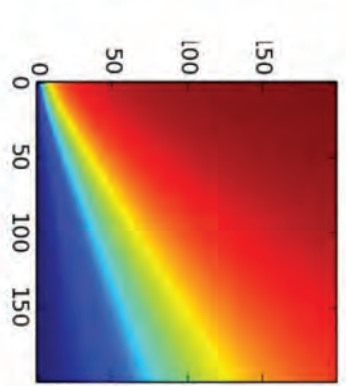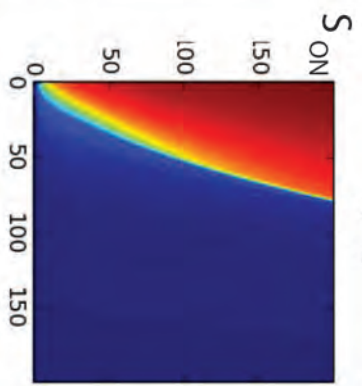
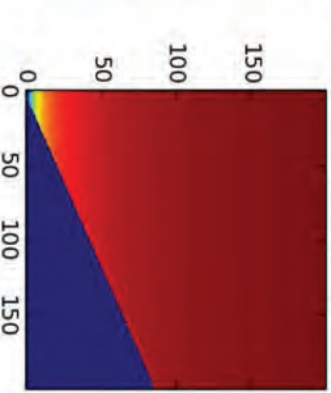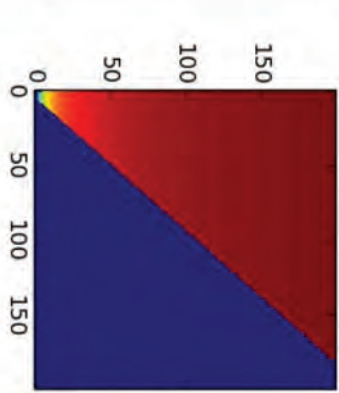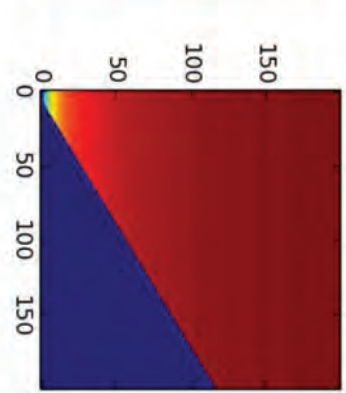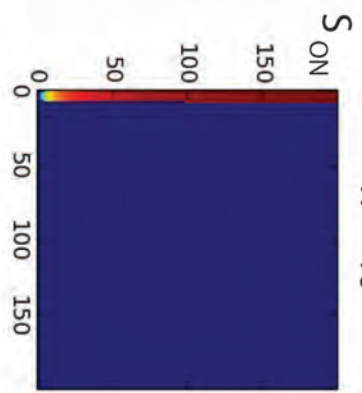

**Figure S10. Phenotype diagrams for a population of *N* cells with a finite Hill coefficient (Strong neighbor communication: *L* > $L_C$; signaling strength $f_N$(L) = 1.5).**

*(Related to Figures 3D and 3E)*

We numerically computed the phenotype diagrams for a population *N* cells with a positive feedback and a Hill coefficient *n* by devising a computational algorithm (see supplementary text). We used four different values for the Hill coefficients: (1) a "low" value (*n*=1), (2) an "intermediate" value (*n*=1.5), (3) a "high" value (*n*=2), and (4) a "near infinite" value (*n*=40). To obtain some intuition for our numerical work, we obtained phenotype diagrams for four initial population states: (1) cell-I is initially OFF and everyone else is initially ON, (2) cell-I is initially OFF and everyone else is initially OFF, (3) cell-I is initially ON and everyone else is initially ON, and (4) cell-I is initially ON and everyone else is initially OFF. To obtain each phenotype diagram, we fixed the Hill coefficient *n* to be one of the four values mentioned above. Then we simulated our computational algorithm for each value of (*K*, $S_{ON}$), which used an iterative algorithm that we designed (we used 50 iterations for each value of (*K*, $S_{ON}$)). Then we used a color to represent a "phenotype score" (see Supplementary text). The phenotype score is necessary because the finite Hill coefficient allows for cells to be in a continuum of states between OFF and ON (instead of the binary ON and OFF states). Doing this for a wide range of values for *K* and $S_{ON}$ resulted in a heat map (phenotype diagram). We used a color spectrum that starts from a pure red (representing a phenotype score of 1) and ends in a pure blue (representing a phenotype score of 0). A pure red represents cell-I being in the ON-state while a pure blue represents cell-I being in the OFF-state. An intermediate color such as green represents the case in which the cell-I is in between the ON and OFF (i.e., partially ON). A color that is closer to red means that cell-I is closer to being ON while a color that is closer to blue means that cell-I is closer to being OFF. Note that the "near infinite" Hill coefficient is nearly identical to the phenotype diagram for cell-I (Figure 3D). Moreover, except for the appearance of intermediate states (e.g., green regions), we find that the main qualitative features of the phenotype diagrams for cells with a finite Hill coefficient are essentially identical to the main features of the phenotype diagrams for cells with an infinite Hill coefficient (Figure 3D).

# Supplemental Theoretical Procedures

## Table of Contents

This document is organized into following sections:

### 1. Three dimensional spherical cell with a finite radius

Instead of treating cells as point objects, we now consider 3-dimensional spherical cells with radius $R$. We consider a 2-dimensional "tissue" formed by a sheet of these spherical cells arranged in a hexagonal lattice, just as in the case of point-like cells. We let $a_o$ be the distance between the centers of two adjacent spherical cells. The $a_o$ is the same lattice constant as in the case of point-like cells. Whereas we considered diffusion equation in two-dimensions for the case of point-like cells, we now consider the diffusion equation in three-dimensions with a constant degradation rate $\gamma$ and a constant secretion rate. First, let's consider a single isolated spherical cell whose center is at $r=0$. The three-dimensional diffusion equation for the isolated cell is

$$\frac{\partial S}{\partial t} = \frac{1}{r^2}\frac{\partial}{\partial r}(Dr^2\frac{\partial S}{\partial r}) - \gamma S + \frac{\eta}{4\pi R^2}\delta(r-R) \tag{1}$$

where $S$ is the concentration outside the spherical cell and $\delta$ is the Dirac delta function. The steady state solution to this equation is spherically symmetric and is

$$S(r) = \frac{S_R R}{r}exp(-\frac{(r-R)}{\lambda}) \tag{2}$$

where

$$S_R \equiv \frac{\eta}{4\pi R^2} \frac{\gamma}{\lambda} \frac{1}{1+\lambda/R} \tag{3}$$

and $\eta$ is the constant secretion rate. $\lambda \equiv \sqrt{D/\gamma}$ as in the case of the point-like cell.

## 1a. Phenotype diagram of the isolated spherical cell

If the cell is "ON", it secretes the signaling molecules at rate $R_{ON}$. The steady state concentration around it becomes

$$S_{ON} = \frac{R_{ON}}{4\pi R^2} \frac{\gamma}{\lambda} \frac{1}{1+\lambda/R} \tag{4}$$

Moreover, the steady state concentration around the OFF cell that secretes the signaling molecule at a constant rate $R_{OFF}$ is

$$S_{OFF} = \frac{R_{OFF}}{4\pi R^2} \frac{\gamma}{\lambda} \frac{1}{1+\lambda/R} \tag{5}$$

Thus we see that $S_{ON}/S_{OFF} = R_{ON}/R_{OFF}$ as in the case of the point-like cell (Fig. S1). That is, the radius of the cell affects both the $S_{ON}$ and $S_{OFF}$ in the same way (Fig. S1). Therefore, for a cell with a given radius $R$, we can measure all concentrations in units of $S_{OFF}$. That is, we can set $S_{OFF} = 1$ as we did for the point-like cell. For this reason, the phenotype diagram for an isolated spherical cell with a given radius $R$, for both the positive and negative feedbacks, is the same as the phenotype diagram of the point-like cell (Figures 2D, 2E, and 2F). Having the cell be spherical instead of being point-like does not change the phenotypes that the cell can have or where the boundaries are between the phenotypes in the phenotype diagram because the radius $R$ scales $S_{ON}$ and $S_{OFF}$ in the same way.

The only difference now is that since $S_{ON}$ and $S_{OFF}$ both depend on the radius radius $R$, the cell can change its radius to decrease both concentrations while keeping the secretion rates $R_{ON}$ and $R_{OFF}$ unchanged (Fig. S1). The only way that a point-like cell can decrease either concentration is by decreasing either the $R_{ON}$ or $R_{OFF}$, whichever is appropriate. As long as the cell tunes its threshold $K$ to match the change in $S_{ON}$ and $S_{OFF}$ associated with the change in radius (i.e., by changing $K$ by a factor $\frac{1}{R^2}\frac{1}{1+\lambda/R}$), the cell would maintain its phenotype after the changing its size.

## 1b. Phenotype diagram of the basic unit composed of spherical cells

Consider the basic hexagonal unit (Fig. 3A) but now with spherical cells with radius $R$. Here we will explicitly treat the scenario in which the spherical cells are close to each other, so that $a_o \sim 2R$. But the basic method that we show below will be the same for spherical cells that are further apart from each other. The only difference would be that the terms that contain distance between cells with look more complicated. We also choose to analyze the case of $a_o \sim 2R$ because in real tissues, cells would be nearly touching each other. Then the concentration $S_I$ that "cell-I" senses is

$$S_I = S_R + \sum_{j=1}^{6} S_{R,j} \frac{R}{r_j} exp(-(r_j - R)/\lambda) \tag{6}$$

where $S_{R,j}$ is the value of the $S_R$ for the j-th cell and $r_j$ is the distance between cell-I and the j-th cell. Since we have $r_j = a_o = 2R$ and $L = \lambda/a_o$, we have

$$S_I = S_R + \sum_{j=1}^{6} \frac{S_{R,j}}{2} exp(-1/2L) \tag{7}$$

From this, we see that we obtain the phenotype diagram as in the point-like cells (Fig. 4A) but now with $exp(-1/2L)$ replacing the $exp(-1/L)$ in the equation for $S_I$ and the $S_{ON}$ now dependent on $R$ (Fig. S1).

### 1c. Phenotype diagram for N spherical cells

We now consider a population of $N$ interacting spherical cells that is assembled by joining multiple basic hexagonal units (Fig. 3A) as we did with the point-like cells. As hinted by our analysis of the basic unit in the previous section, we will get a phenotype diagram that is qualitatively the same as that of the population of $N$ point-like cells (Fig. 4C) but now with a slightly different equation for the straight lines that separate the different phenotype regions (Fig. 4C). Recall from the main text that these boundary lines are defined by the setting the phenotype function to zero: $\varphi(K, S_{ON}) = 0$. First, cell-I now senses the following concentration:

$$S_I = S_R + \sum_{j=1}^{N-1} S_{R,j} \frac{R}{r_j} exp(-(r_j - R)/\lambda) \tag{8}$$

To construct the phenotype diagram for $N$ cells, we need to solve the equation $S_I - K = 0$ with the 4 limiting cases for $(C, \Omega)$ as we did for the point-like cells. For example, in the limiting case in which the cell-I and everyone else in the population are OFF, we obtain (with the signaling length $L = \lambda/a_o$ defined in the same way as in the point-like cells),

$$S_I = 1 + R \cdot exp(R/L) \sum_{j=1}^{N-1} \frac{1}{r_j} exp(-r_j/L) \tag{9}$$

where all lengths (R and $r_j$'s) are measured in units of the lattice constant $a_o$ as we did in the case of point-like cells. This looks very similar to the equation we get for point-like cells, except now it depends on the cells' radius $R$.

From above example, we see that we can define the signaling strength function as we did in the point-like cells by looking the summation in above equation. The form would now look different, and importantly, it now depends on the cell radius $R$. Specifically, the signaling strength function $f_{N,R}(L)$ for the spherical cells is

$$f_{N,R}(L) \equiv R \cdot exp(R/L) \sum_{j=1}^{N-1} \frac{1}{r_j} exp(-r_j/L) \tag{10}$$

We note that this is very similar to the signaling strength function $f_N(L)$ of the $N$ point-like cells. The main difference is that now this function depends on $1/r_j$ and the radius $R$. This formula is general and thus holds for any separation distance $a_o$ (not just for spherical cells nearly touching each other) (Fig. S1).

## 2. Entropy of population

Before trying to find the "equation of motion" that describes how a population state $(C, \Omega)$ of $N$ secrete-and-sense cells evolves over time, we can consider two cases: (1) A population is stable thus communications among cells will not change its state $(C, \Omega)$ over time or (2) A population's state change over time due to cell-cell communication. We say that the population in the first scenario is in equilibrium. We call its state an "equilibrium state". Finding the total number of equilibrium states is our first step towards obtaining the equation of motion for the population state. Below we derive a formula that estimates the number of equilibrium states.

### 2a - Derivation of equation (8) in the main text

We can express the total number $\Omega_E$ of equilibrium populations as a sum of the number $\Omega_k$ of equilibrium populations in which $k$ cells are ON, for each possible value of $k$ : $\Omega_E = \sum_{k=0}^{N} \Omega_k$

For a population of $N$ cells and $k$ ON cells, we have $\binom{N}{k}$ possible states, so that $\Omega_k = p_e(k, N) \cdot \binom{N}{k}$, with $p_e(k, N)$ the fraction of equilibria. In total we get :

$$\Omega_E = \sum_{k=0}^{N} p_e(k, N) * \binom{N}{k} \tag{11}$$

We can find an estimation $\tilde{p}_e$ of $p_e$ by considering a population with $k$ ON cells as a random variable. Each cell of such a population is a random variable, can be ON with probability $p = k/N$ and OFF with probability $1 - p$. By that means we can estimate $p_e$ by the probability of a random population to be an equilibrium.

Let us consider a population of $N$ cells that has $k$ ON cells. Let $p$ be the fraction of ON cells, $p \equiv k/N$. We consider each cell's state $C_i$ to be a random variable that follows the Bernoulli distribution as a function of $p$. Note that $C_i = 1$ if the cell is ON and $C_i = 0$ if the cell is off. Moreover each $C_i$ is an independent variable that follows the same Bernoulli distribution as all the other cell states. A population of cells is then defined by an N-dimensional random variable, $Z_N \equiv (C_1, C_2, ..., C_N)$. Since each cell state is a random variable, the steady state concentration on the cell surface is also a random variable that is correlated with the cell state. We define this random variable $X_i$, and it is

$$X_i = S_{ON} * C_i + (1 - C_i) \tag{12}$$

Here we used the fact that we are working in units of concentration in which $S_{OFF} = 1$. The total concentration $Y_i$ that a cell-i senses, which is due to the signalling molecules from all $N$ cells, is also a random variable:

$$Y_i = X_i + \sum_{j \neq i} X_j \cdot e^{-\frac{r_{ij}}{L}}, \qquad (13)$$

This is just the equation (6) from the main text but now recast in terms of the random variable $X_j$. We want to calculate the probability that $Z_N(\omega)$ is an equilibrium state. We let $p_e(p, N)$ be the probability that a population of $N$ cells with a fraction $p$ of ON cells is in equilibrium. For cells with the positive feedback (Fig. 1d in main manuscript), being in equilibrium means that for any ON cell-i we have $Y_i > K$ whereas for any OFF cell-j we have $Y_j < K"$. A population is in equilibrium if and only if both conditions are satisfied for every cell in the population. Stating this in set theory notation, we have

$$\{1, 2, ..., N\} = [\{i : C_i = 0\} \cap \{i : Y_i < K\}] \cup [\{i : C_i = 1\} \cap \{i : Y_i \geq K\}] \qquad (14)$$

and since it has to be true for every cell, we have

$$p_e(p, N) = \bigcap_{i=1}^{N} [[(C_i = 0) \cap (Y_i < K)] \cup [(C_i = 1) \cap (Y_i >= K)]] \qquad (15)$$

The difficulty that we face here is that not all $Y_i$'s are independent of each other because some cells share the all same $C_i$s. Thus analytically exacting the value of $p_e(p, N)$ is difficult. Instead of trying to find its exact value, let's assume that the $Y_i$'s are weakly dependent (i.e., almost independent) of each other, due to the exponentially decaying value of the concentration as a function of distance from the secreting cell. We will check the validity of this assumption later by checking the formula that this assumption leads us to with the results of exact simulations. Treating all $Y_i$s to be independent of each other, we have

$$p_e(p, N) = \prod_{i=1}^{N} P([[(C_i = 0) \cap (Y_i < K)] \cup [(C_i = 1) \cap (Y_i \geq K)]]) \qquad (16)$$

where $P(S)$ is the probability that statement $S$ is satisfied. Since each $C_i$ follows the same Bernoulli distribution, we have

$$p_e(p, N) = P([[(C_i = 0) \cap (Y_i < K)] \cup [(C_i = 1) \cap (Y_i \geq K)]])^N \qquad (17)$$

A cell cannot be simultaneously ON and OFF. Thus above equation becomes

$$P([(C_i = 0) \cap (Y_i < K)] \cup [(C_i = 1) \cap (Y_i \geq K)]]) = P([[(C_i = 0) \cap (Y_i < K)] + P([(C_i = 1) \cap (Y_i \geq K)]]) \qquad (18)$$

Moreover, we note that

$$P([[(C_i = 0) \cap (Y_i < K)]) = P(C_i = 0) * P_{C_i=0}(Y_i < K) = (1-p) * P_{C_i=0}(Y_i < K), \qquad (19)$$

where $P_{C_i=0}$ is the conditional probability given that $C_i = 0$. Substituting this result into equation (17), we obtain

$$p_e(p, N) = [(1-p) * P_{C_i=0}(Y_i < K) + p * P_{C_i=1}(Y_i > K)]^N \tag{20}$$

Now let's find the two conditional probabilities, $P_{C_i=0}(Y_i < K)$ and $P_{C_i=1}(Y_i > K)$. Finding their exact expressions is challenging. But note that $Y_i$ is a sum of a large number ($N$) of the independent terms $X_i e^{-\frac{r_{ij}}{L}}$, each of which is small due to the exponential decay term. Thus by an extension of the central limit theorem, we assume that each $Y_i$ follows a Gaussian distribution. More precisely, we assume that each $Y_i$ satisfies the Lyapunov's condition on moments thus it converges to the normal distribution. With this assumption, our problem is reduced to finding the mean and the standard deviation of $Y_i$, conditional on $C_i = 0$ or $C_i = 1$. We find that

$$<Y_i>_{(C_i=0)} = <X_i + \sum_{j \neq i} X_j \cdot e^{-\frac{r_{ij}}{L}}>_{(C_i=0)} = <1 + \sum_{j \neq i} X_j \cdot e^{-\frac{r_{ij}}{L}}> \tag{21}$$

where we use the notation $<\cdot>_{(C_i=0)}$ to denote the mean value of $\cdot$ conditional on $C_i = 0$. Now since we have

$$<X_j> \sum_{j \neq i} e^{-\frac{r_{ij}}{L}} = [S_{ON} * p + (1-p)] * f_N(L), \tag{22}$$

equation (21) becomes

$$<Y_i>_{(C_i=0)} = 1 + [\alpha_{ON} * p + (1-p)] * f_N(L) \tag{23}$$

and similarly

$$<Y_i>_{(C_i=1)} = S_{ON} + [S_{ON} * p + (1-p)] * f_N(L) \tag{24}$$

Next we compute the variance of $Y_i$ conditional on a given state $C_i$. We use $\text{Var}(\cdot)_{(C_i=0)}$ to denote the variance of $\cdot$ given that $C_i = 0$. We then have

$$Var(Y_i)_{(C_i=0)} = Var(X_j) \sum_{j \neq i} (e^{-\frac{r_{ij}}{L}})^2 \tag{25}$$

Since all $X_j$'s are mutually independent of each other, above equation becomes

$$Var(Y_i)_{(C_i=0)} = (1-p) * p * (S_{ON} - 1)^2 * \sum_{j \neq i} (e^{-\frac{r_{ij}}{L}})^2 \tag{26}$$

Repeating the calculation for $Var(Y_i)_{(C_i=1)}$, we find that $Var(Y_i)_{(C_i=1)} = Var(Y_i)_{(C_i=0)}$.

Given that the mean and the variance of $Y_i$ does not depend on the $i$ under the periodic boundary condition that we use throughout our work (i.e., under the periodic boundary condition, $f_N(L)$ is purely a geometric property of the lattice on which the cells are placed), we can define $\mu_{OFF,p} \equiv <Y_i>_{C_i=0}$, $\mu_{ON,p} \equiv <Y_i>_{C_i=1}$, and $\sigma_p^2 \equiv Var(Y_i)_{(C_i=0)}$. In summary, we have

$$\begin{aligned}
\mu_{OFF,p} &= 1 + [S_{ON} * p + (1-p)] * f_N(L) \\
\mu_{ON,p} &= S_{ON} + [S_{ON} * p + (1-p)] * f_N(L) \\
\sigma_p^2 &= (1-p) * p * (S_{ON} - 1)^2 * \left(\sum_{i=1}^{N-1} (e^{-\frac{r_i}{L}})^2\right)
\end{aligned} \quad (27)$$

We are now ready to estimate the $p_e(p,N)$. For the reasons mentioned before, we can invoke the central limit theorem to get

$$\begin{aligned}
P_{C_i=0}(Y_i < K) &\sim \phi(\frac{K - \mu_{OFF,p}}{\sigma_p}) \\
P_{C_i=1}(Y_i > K) &\sim 1 - \phi(\frac{K - \mu_{ON,p}}{\sigma_p})
\end{aligned} \quad (28)$$

where $\phi$ is the cumulative distribution function for the normal distribution with a mean of 0 and a standard deviation of 1. Substituting these into equation (20), we have

$$p_e(p, N) \approx [(1-p) * \phi(\frac{K - \mu_{OFF,p}}{\sigma_p}) + p * (1 - \phi(\frac{K - \mu_{ON,p}}{\sigma_p}))]^N \quad (29)$$

Thus the total number $\Omega_E$ of populations that are in equilibrium for a given $(K, S_{ON}, L)$ is

$$\Omega_E \sim \sum_{k=0}^{N} [(1-p) * \phi(\frac{K - \mu_{OFF,p}}{\sigma_p}) + p * (1 - \phi(\frac{K - \mu_{ON,p}}{\sigma_p}))]^N * \binom{N}{k} \quad (30)$$

And as we did in the main text, we define the entropy of population $\sigma$ as

$$\sigma = \frac{\Omega_E}{2^N} \quad (31)$$

### 2b. - Comparison with simulation

Now we can check how closely our estimation for the entropy of population $\sigma$ (equation 31) matches the its true value. To do so, we empirically obtained $\sigma$ from exact computer simulations of population dynamics. For each value of $(K, S_{ON}, L)$ and $k$ ($0 \leq k \leq N$), we performed one thousand simulations. In each simulation, we randomly a state $C_i$ to each one of the $N$ cells by using the Bernoulli distribution with $p = \frac{k}{N}$. We then compute the concentration $Y_i$ (equation (13)) for each cell. This determines if any cell's state needs to change. If no cell's state changes, then we count this population state to be an equilibrium state. We then repeat the simulation by randomly picking the initial cell states again. Doing this 1000 times, we obtain an empirical estimate of the $p_e(p, N)$. Doing this for a wide range of values for $(K, S_{ON}, L)$, we find that our estimate (equation 31) closely matches the value that we obtain through our exact simulations (e.g., see Fig. 5a in the main text).

### 3. Model for population dynamics without spatial arrangements of cells

Our results in the main text suggest that spatial clustering of cells can strongly influence how the ON/OFF state of each cell in a population changes over time. To quantify this effect, we constructed analytical model in which N cells are in a uniformly mixing liquid culture (thus no spatial arrangements). This "mean-field" model describes how the number of ON cells in this culture changes over time due to cell-cell signaling. Since the cells would move in the liquid culture, it will randomly encounter ON and OFF cells. This is equivalent to considering the spatially fixed population of cells (Figure 3A) but now with the spatial clustering index IM equal to zero (i.e., ON and OFF cells are randomly arranged in space). The mean-field model also treats scenarios in which the cells make errors in their secretion or sensing. For different values of initial fraction of ON cells, we used the mean-field model to compute the temporal change in the number of ON cells and compared it with the temporal change predicted by the exact simulations (Figure S7). We found that the difference between the two temporal trajectories of the fraction of ON cells increased as the clustering index IM increased (i.e., as cells became more spatially ordered). In this way, our mean-field model directly quantified the effects of spatial clustering of cells on the population dynamics.

Below we derive a mathematical model to describe the temporal evolution of any population state $(C, \Omega)$ without taking into account the spatial arrangements of cells. This "mean-field" model does not take into account all the spatial details of the population. After deriving the "equation of motion" that describes how the fraction $p$ of ON cells changes over time in our mean-field model, we will compare it with the temporal change in $p$ that we obtain from exact simulations that does account for the spatial location of each cell.

### 3a. Derivation of the mean-field model: Discrete version

A population of $N$ cells can be in any one of $2^N$ possible states. Thus if we apply our deterministic model of cellular communication (equation (6) $S_I$ in the main text), we would obtain the exact relationship between each of the $2^N$ states (i.e., given a population with state $(C_i, \Omega)$, we would know which one of the other states $(C_j, \Omega')$ the population would enter at the next time step). So we can, in principle, build a "network diagram" in which the nodes are each of the $2^N$ states and directed arrows between them represent a transition between the states governed by the equation (6) in the main text. But there are too many states for this approach to be practical. Importantly, we wouldn't necessarily obtain a sense that we understood the important aspect of the dynamics. In order to simplify the problem, we can consider classes of population states instead of individual population states. To do so, we group different population states together into one class if they have the same fraction $p$ of ON cells in them. We have exactly $\binom{N}{pN}$ states that have a fraction of ON cells $p$. There are $N+1$ such classes of states. Let us now deduce the transition probabilities between any pair of classes. Suppose we take a random population state with a fraction $p_j$ of ON cells. Our main idea is that we can then compute $S_I$ for each cell (equation (6) in the main text) in this state and then deduce the probability $P_{p_j}(p_i)$ that the obtained state has a fraction $p_i$ of ON cells.

With the same notations and logic used in the previous section, we can approximate $P_{p_j}(p_i)$ by the binomial distribution. Thus we have

$$P_{p_j}(p_i) = \binom{N}{Np_i} \alpha^{p_i N}(1-\alpha)^{(1-p_i)N} \tag{32}$$

with $\alpha \equiv 1 - (1-p_j) \cdot \phi(\frac{K-\mu_{OFF,p_j}}{\sigma_{p_j}}) - p_j \cdot \phi(\frac{K-\mu_{ON,p_j}}{\sigma_{p_j}})$. There are $N+1$ possible values of $p$, and thus we have $(N+1)^2$ values of $P_{p_j}(p_i)$. We define a N+1 by N+1 transition matrix $M = (a_{ij})$ where $a_{ij} \equiv P_{p_j}(p_i)$. The transition matrix $M$ is a function of $S_{ON}$, $K$ and $L$, and represent a statistical approximation to how each population transitions into a different state. At time $t$ we have a population with a fraction $p(t)$ of ON cells, the state at the next time step $p(t+1)$ is $<p>_{p(t)}$. Due to the binomial distribution, we have $<p>_{p(t)} = N \cdot \alpha$. By repeating this process, we can then estimate the evolution over time of $p$. An important remark is that this model does not take into account any spatial arrangements of cell. Instead it is the minimal model that we would have if the population was well mixed in space because we assumed a randomly distributed cells in a population for each $p$.

### 3b. Derivation of the mean-field model: Continuum version

We now derive a mean-field model that treats time $t$ to be continuous rather than as a discrete as we did in the previous section. To get a mean-field model, we consider the concentration of the signalling molecule to be uniformly spread out in the population. This is equivalent to the statement that we pick a random cell within a population, then it's likely to be any one of the cell's in the population. Thus the concentration sensed by this randomly selected cell would be the $S_I$ (equation (6)) averaged over all the cells. Using the same notations as before, the average concentration $<S_{neighbours}>_p$ of the molecules from only the neighbouring cells is

$$
\begin{aligned}
<S_{neighbours}>_p &= <\tfrac{1}{N} \cdot \sum_{i=1}^{N}(\sum_{j \neq i} \alpha_{Oj} \cdot e^{-\tfrac{r_{ij}}{L}})>_p \\
&= (p \cdot S_{ON} + (1-p)) \cdot f_N(L)
\end{aligned}
\tag{33}
$$

where we have used the same calculations in deriving equation (22) in the previous section. Then the mean value of the concentration $S_I$ sensed by a randomly chosen cell (cell-I) in the population is

$$
\begin{aligned}
<S_{I_{OFF}}>_p &= 1 + (p \cdot S_{ON} + (1-p)) \cdot f_N(L) \\
<S_{I_{ON}}>_p &= S_{ON} + (p \cdot S_{ON} + (1-p)) \cdot f_N(L)
\end{aligned}
\tag{34}
$$

where $<S_{I_{OFF}}>_p$ and $<S_{I_{ON}}>_p$ are the average concentration sensed by an OFF cell and an ON cell respectively. Now we propose an "equation of motion" by using $<S_{I_{OFF}}>_p$ and $<S_{I_{ON}}>_p$ to mimic the step function (Figs. 1D and 1E in the main text) that represents the secretion rate as a function of the concentration of the sensed molecule. Specifically, we impose the probability $P_{OFF \to ON}$ that a cell transitions from OFF to ON and the probability $P_{ON \to OFF}$ that a cell transitions from ON to OFF to be sigmoidal functions of $<S_{I_{OFF}}>_p$ and $<S_{I_{ON}}>_p$ respectively. In the case of cells with the positive feedback, we let

$$
\begin{aligned}
P_{OFF \to ON} &= \frac{1}{1+(\frac{K}{S_{I_{OFF}}})^c} \\
P_{ON \to OFF} &= \frac{1}{1+(\frac{S_{I_{ON}}}{K})^c}
\end{aligned}
\tag{35}
$$

where $c$ is a hill coefficient. Based on these "transition functions" $P_{OFF \to ON}$ and $P_{ON \to OFF}$, which are continuous versions of the discrete transition matrices we defined above, we obtain the following "equation of motion" that describes how a population with $p$ ON cells evolves over time:

$$\frac{dp}{dt} = p \cdot (P_{ON \to OFF} + \eta) + (1-p) \cdot (P_{OFF \to ON} + \eta) \tag{36}$$

where we have added the noise term $\eta$ to allow for stochastic effects in cell-cell signalling (e.g., cell makes a mistake in sensing and secretion near the threshold $K$ due to the sharpness of the sigmoidal function). By rearranging the terms in above equation, we can rewrite it as

$$\frac{dp}{dt} = \underbrace{\eta(1-2p)}_{Noise} + \underbrace{p \cdot P_{ON \to OFF} + (1-p) \cdot P_{OFF \to ON}}_{signalling} \tag{37}$$

Above is the continuous mean-field model's "equation of motion".

## 4. Remarks on the clustering index - A measure of spatial arrangement of cell states (equation (11) in the main text)

Here we motivate and give an intuitive idea behind Moran's I (our "clustering index"), defined in equation (11) in the main text. In order to quantify the spatial arrangement ( clustering ) of cells on a lattice, our clustering index (Moran's I with our own definition of the "weight" term $w_{ij}$) compares two statistical quantities:

- Spatial weighted covariance: $Cov_s = \frac{1}{\sum_{i=1}^{N} \sum_{j=1}^{N} w_{ij}} \sum_{i=1}^{N} \sum_{j=1}^{N} w_{ij}(C_i - \bar{C})(C_j - \bar{C})$, with weight $w_{ij}$ between $C_i$ and $C_j$ defined as : $w_{ij} \equiv \frac{1}{r_{ij}}$, where $r_{ij}$ is the distance between cell-i and cell-j. Here we denote $\bar{C} \equiv \frac{1}{N} \sum_{i=1}^{N} C_i$)

- Variance $V = \frac{1}{N} \sum_{i=1}^{N} (C_i - \bar{C})^2$

Our clustering index is then defined to be

$$I_M \equiv \frac{Cov_s}{V} = \left[\frac{1}{\sum_{i=1}^{N} \sum_{j=1}^{N} w_{ij}} \sum_{i=1}^{N} \sum_{j=1}^{N} w_{ij}(C_i - \bar{C})(C_j - \bar{C})\right] \cdot \frac{N}{\sum_{i=1}^{N} (C_i - \bar{C})^2} \tag{38}$$

which is just the ratio between the $Cov_s$ and the $V$. Note that $-1 \leq I_M \leq 1$, with $I_M = 0$ when cell states are randomly arranged on a lattice and $I_M = 1$ when there is a perfect clustering (i.e., all ON cells are clustered together in one region). We can intuitively understand $I_M$ in the following way. Consider a regional cluster of cells that are in with same state. Then their cross product is $(C_i - \bar{C})(C_j - \bar{C}) = (C_i - \bar{C})^2$. Furthermore, say the $r_{ij}$ between any two cells in this cluster is low so that $w_{ij} \sim 1$. Then for this cluster, we have $w_{ij}(C_i - \bar{C})(C_j - \bar{C}) \sim (C_i - \bar{C})^2$. We clearly see here that the more the population is clustered the more $\frac{1}{\sum_{i=1}^{N} \sum_{j=1}^{N} w_{ij}} \sum_{i=1}^{N} \sum_{j=1}^{N} w_{ij}(C_i - \bar{C})(C_j - \bar{C})$ gets close to $\frac{1}{N} \sum_{i=1}^{N} (C_i - \bar{C})^2$. In addition, $I_M$ increases towards 1.

In our work, although we have used the periodic (toric) boundary condition to compute the concentration of the signaling molecule, we computed the clustering index by treating the edges of the field of cells as hard (non-periodic) boundaries. This simplified our calculations without affecting the main qualitative conclusions about spatial ordering of cells. With the periodic boundary conditions, there are two possible separation distances between any pair of cells: (1) Short distance and (2) long distance (e.g., distance measured between two cells by traversing through left and right edges of the lattice that are joined together). By using hard boundaries, every pair of cells has one separation distance between them.

## 5. Remarks on the neighbour-induced activation, neighbour-induced deactivation, and the neighbour-induced activation-deactivation phenotypes

Here we describe in more detail the collective phenotypes: neighbour-induced activation, neighbour-induced deactivation, and neighbour-induced activation-deactivation (regions of the phenotype diagrams shown in Fig. 4C in the main text).

The neighbour-induced activation region (green region in Fig. 4C in the main text): An OFF cell-I in this region can potentially be turned ON if a sufficiently high density of neighbouring cells are ON. By density, we mean that the combination of the number and location of ON neighbouring cells is the deciding factor in whether or not an OFF cell-I can be activated into an ON state. Since the concentration of the secreted signalling molecule decreases with distance from the secreting cell, a large number of distant ON cells and a small number of nearby ON cells can both produce the same concentration of the signalling molecule around cell-I. If cell-I is ON, it will stay ON whatever the state of the other cells are because the concentration of the signal produced by the cell-I itself is sufficient to keep it ON.

The neighbour-induced deactivation region (brown region in Fig. 4c in the main text): We have the exact opposite behaviour from that of the "active region". If cell-I is OFF, it will stay OFF regardless of the number of OFF and ON cells in the rest of the population. If cell-I is ON, it will stay ON only if a sufficiently high number of the neighbours are ON. Otherwise cell-I will turn OFF.

In the neighbour-induced activation-deactivation region (white region in Fig 4c - right panel, in the main text), we have a simultaneous existence of activation and deactivation, each of which we separately described above. Here, if cell-I is ON it can be deactivated if a sufficiently high number of cells are OFF. If the cell is OFF, it can be activated if the ON cells in the population are sufficiently close to cell-I or present in large numbers. Thus the entire population is sensitive to both an increase and a decrease of the number of ON cells. Moreover the population state is also sensitive to the spatial arrangements of ON and OFF cells.

## 6. Finite Hill coefficient

In the main text, we used step functions to approximate the positive and negative feedback regulations (Figures 1D and 1E). In other words, we assumed that the secretion rate of the signaling molecule was described by a sigmoidal function of the sensed concentration with an infinite Hill coefficient. Although this idealisation simplified our calculations and has been shown to capture essential features of several important gene regulatory systems, it is important to ask how using a finite Hill coefficient would affect the main results that we obtained in the main text. For instance, we are interested in understanding how the phenotype diagrams that we obtained from using the step functions in the main text (Figs. 2-4 in the main text) would change if we use a finite Hill coefficient. We now address this question.

Here we restrict our attention to cells with the positive feedback because the negative feedback follows the same principle. Consider a cell with the positive feedback and a finite Hill coefficient $n$. Before delving into any calculations, let us first intuitively see how such a cell would regulate its secretion rate. Suppose that a cell is initially sensing some concentration $S_1$ and secretes the signalling molecule at some rate $F(S_1)$ in response to it. This secretion causes the cell to establish a new concentration on its surface. Namely, if the secretion rate remains constant at $F(S_1)$, then the cell would establish a new steady state concentration $S_2$ on its surface after some time. This concentration would be directly proportional to the rate $F(S_1)$. This is true for both point-like and spherical cells (see equation (3)). Sensing this new concentration on its surface, the cell would then readjust its secretion rate to $F(S_2)$. After some time, the establishes a new steady-state concentration $S_3$, which is proportional to the secretion rate $F(S_2)$. The cell then changes it secretion rate to $F(S_3)$. To treat this scenario, we can consider a small, fixed discrete time step $\delta t$. The $\delta t$ represents the time that the cell requires to measure and respond to the the concentration of the signalling molecule outside. No cell can instantaneously measure the concentration outside. Cells, even those that are not secrete-and-sense cells, would take the average of many measurements of the concentration made over some time (known as the "integration time", *a la* Berg and Purcell) and set this average as *the* concentration outside it. Indeed this is what one means by saying that the secretion rate is a sigmoidal function (or any other function) of the concentration sensed by the cell. It is also for this reason that we can think in terms of a series of discrete time steps of interval $\delta t$ during which the secretion rate is held constant. Our main idea from here on is to quantitatively describe the sequence of events described above, in which the cell iteratively readjusts its secretion rate as a function of the quasi-statically changing concentration outside it. We will then see if he cell converges to one or more possible equilibrium states that are analogous to the "ON" and "OFF" states of the cells with the infinite Hill coefficient.

Let $F_n(S_I)$ be the rate at which the cell secretes the signaling molecule when it senses concentration $S_I$. Then we note that for a positive feedback regulation, $F_n$ is a sigmoidal function of $S_I$ with a Hill coefficient $n$. Namely,

$$F_n(S_I) = R_{OFF} + \frac{(R_{ON} - R_{OFF}) \cdot S_I^n}{K^n + S_I^n} \qquad (39)$$

is the secretion rate when the cell detects concentration $S_I$ on its surface. Note that since the cell secretes a continuum of values between $R_{OFF}$ and $R_{ON}$, we cannot say whether a cell is "ON" or "OFF". The cell secretes at a constant rate $F_n(S_I)$ during the integration time interval $\delta t$ because the cell takes the time $\delta t$ to change its gene expression, including the expression level of the gene that encodes the signaling molecule (Fig. 1C). We denote the resulting concentration on the cell after this time as $f_n(S_I)$. Note that for both point-like and spherical cells, this concentration is proportional to the secretion rate. Thus multiplying $F_n(S_I)$ by a constant factor, we obtain $f_n(S_I)$. Namely,

$$f_n(S_I) = S_{OFF} + \frac{(S_{ON} - S_{OFF}) \cdot S_I^n}{K^n + S_I^n} \qquad (40)$$

is the newly established concentration, after an integration time $\delta t$, on a cell that initially senses concentration $S_I$ (Fig. S8). Note that when the Hill coefficient is infinite, we have

$$f_\infty(S_I) = \begin{cases} S_{OFF} & \text{if } S_I < K \\ S_{ON} & \text{if } S_I \geq K \end{cases} \qquad (41)$$

This matches the step-function regulation scheme that we analyzed in the main text. To be concrete, note that if the cell senses concentration $S_1$, then it would create concentration $f_n(S_1) = S_2$ after time $\delta t$. The cell would adjust its secretion rate to $F_n(S_2)$. This results in a new concentration $f_n(S_2) = S_3$ after time $\delta t$. And this sequence would continue. We are interested in whether this sequence eventually converges to an equilibrium and if so, to what value.

### 6a. Isolated cell

Let us consider an isolated cell, which can be either point-like or spherical. It uses the positive feedback regulation with a Hill coefficient $n$. Suppose that the cell initially senses concentration $S_1$. After time step *deltat*, the cell will have created a concentration $f_n(S_1) = S_2$. The cell will then sense this new concentration, readjust its secretion rate so that after time $\delta t$, it establishes a new concentration $f_n(S_2) = S_3$. We see that the sequence of concentrations that are established on the cell surface is

$$S_{t+1} = f_n(S_t) \qquad (42)$$

where we have set the characteristic time step $\delta t$ to be equal to 1 without loss of generality. By looking at the sign of $S_{t+1} - S_t = f_n(S_t) - S_t$, we can determine whether the sequence is increasing (i.e., $f_n(S_t) > S_t$), decreasing (i.e., $f_n(S_t) < S_t$), or if it remains fixed (i.e., $f_n(S_t) = S_t$). This allows us to classify the dynamics into 3 scenarios. They are: (1) self activation, (2) self deactivation, and (3) bistability. We can graphically see these three scenarios (Fig. S8). These three scenarios are analogous to the three phenotypes, "ON", "OFF", and "bistability", of the cells that have an infinite Hill coefficient (Equation [3] in the main text).

The main difference between cells with a finite Hill coefficient and the cells with an infinite Hill coefficient is that the cell takes a longer time (i.e., more time steps) to reach an equilibrium (i.e., ON or OFF state ). Our analysis shows that the higher the Hill coefficient, the faster the cell reaches its stable equilibrium state. As we graphically show (Fig. S8), this makes intuitive sense because the lower the Hill coefficient is, the slower the cell would "climb up" or "slide down" the sigmoidal curve.

## 6b. Population of N cells with a finite Hill coefficient

We now consider a population of $N$ cells with a positive feedback and a finite Hill coefficient. As in the main text, we arrange the $N$ cells in a regular polygonal lattice. Let us designate a particular cell as "cell-I" as in the main text. Following the quasi steady-state approach that we introduced above, let $S_O$ denote the concentration created by cell-I on its surface after detecting concentration $S_I$. As in the case of the isolated cell, the "output concentration" $S_O$ is a sigmoidal function of the "input concentration" $S_I$. Namely,

$$S_O \equiv g_{n,K,S_{ON}}(S_I) = S_{OFF} + \frac{(S_{ON} - S_{OFF}) \cdot S_I^n}{K^n + S_I^n} \qquad (43)$$

By measuring all concentrations in units in which $S_{OFF} = 1$ as in the main text, we have

$$S_O = g_{n,K,S_{ON}}(S_I) = 1 + \frac{(S_{ON} - 1) \cdot S_I^n}{K^n + S_I^n} \qquad (44)$$

In addition, the total concentration that cell-I senses is the sum of the concentration that it creates on itself and the concentration that all the other cells generate on cell-I's. Thus we have

$$S_I = S_O + \sum_{j=1}^{N-1} S_{Oj} exp(-(r_j/L)) \qquad (45)$$

Here $S_O$ is the concentration created by cell-I on itself and $S_{O,j}$ is the $S_O$ for j-th cell. If the Hill coefficient $n$ is infinite, then $S_O$ can take on only one of two values, $S_{ON}$ or 1. This is the scenario that we treat in the main text. For a finite Hill coefficient, $S_O$ can take on a continuum of values between 1 and $S_{ON}$. Hence cell-I can potentially be in an infinite number of secretion states. However, to make progress analytically and to derive results that we can compare with those of the cell with an infinite Hill coefficient, we define a cell to be "ON" if $S_O$ is sufficiently "close to" $S_{ON}$ and OFF if it's sufficiently close to 1. To construct phenotype diagrams and have some information on the population-level behaviors, we would like to know how cell-I would behave over time if cell-I is initially in a state $C$ (Fig. 3C) (i.e., $S_O$ close to $S_{ON}$ or 1 ) and the rest of the population is initially in neighbor-state $\Omega$ (defined in Fig. 3C). To properly define the neighbor state $\Omega$, we denote any neighboring cell whose secretion rate is sufficiently close to the maximal secretion rate $R_{ON}$ to be in an "ON" state and any neighboring cell whose secretion rate is sufficiently close to the minimal secretion rate $R_{OFF}$ to be in an "OFF" state, just as we do for denoting cell-I's state. Thus the population-state (C, $\Omega$) is well defined even with the finite Hill coefficient.

When we assumed that cell-I had an infinite Hill coefficient, we saw that cell-I could either stay in the same initial state or switch to the other state (i.e., ON to OFF or OFF to ON ). But now this is no longer true because $S_O$ can take many different values over time. In fact the sequence of values taken on by $S_O$ can be obtained by solving the coupled equations (44) and (45) with a given initial population-state $(C, \Omega)$. Specifically, we obtain a recursive relation for $S_O$ over time. If we let $S_{O,t}$ denote the value of $S_O$ after time $t$ (where $t$ is an integer because we are taking discrete time steps of size $\delta t = 1$ as in our analysis of the isolated cell), then we have

$$S_{O,t+1} = g_{n,K,S_{ON}}(S_{O,t} + \sum_{j=1}^{N-1} S_{Oj} exp(-(r_j/L))) \qquad (46)$$

With this recursion relation, we can now investigate if cell-I's state (i.e., the value of $S_{O,t}$) converges to an equilibrium state after a long time. Moreover, if the state does converge to an equilibrium state, then we can ascertain if $S_{O,t}$ converges to $S_{ON}$ or 1. By answering this question, we can obtain a comprehensive picture of a population of $N$ cells with a finite Hill coefficient. Namely, we would be able to predict for a given set of values for $(n,K,S_{ON})$ and a given initial neighbor-state $\Omega$, which state cell-I would converge to. This yields a phenotype diagram for cells with a finite Hill coefficient.

Graphically, we can see that when cells have the positive feedback with a finite Hill coefficient, we always get either one or two stable equilibriums (Fig. S8). Moreover we see that there are 3 possible scenarios: (1) Activation, (2) De-activation, and (3) bistability (Fig. S8). Now we would like to know in which of these three scenarios cell-I falls under for a given set of values for $(n, K, S_{ON})$. Unfortunately one cannot analytically solve the recursion relation (46). But we can numerically solve it with the following algorithm, and thus numerically compute the phenotype diagram for $N$ cells with a finite Hill coefficient $n$:

**Computational algorithm to compute the phenotype diagram for $N$ cells with a finite Hill coefficient:**
**Step 1**. For a given set of values for $(n, K, S_{ON})$ and an an initial population state $(C,\Omega)$, we compute the value of $S_{O,t}$ for a sufficiently large $t$. We pick a large $t$ because we observed from our simulations that $S_O$ rapidly converges to an equilibrium.
**Step 2**. We then assign a "phenotype score" whose values is between 0 and 1. The phenotype score is a measure of how close the final value of $S_O$ is to $S_{ON}$ and $S_{OFF} = 1$. If the final value of $S_O$ is closer to $S_{ON}$, then the phenotype score is closer to 1. If the final value of $S_O$ is closer to $S_{OFF} = 1$, then the phenotype score is closer to 0.

We ran this algorithm for two regimes of the signaling length ($L < L_c$ and $L > L_c$). We studied four different values of the Hill coefficient $n$: (1) a "low" value ($n = 1$), (2) an "intermediate" value ($n = 1.5$), (3) a "high" value ($n = 2$), and (4) a "near infinite" value ($n = 40$). We note that $n=40$ already nearly matches a step function (Fig. S8) and that $n=2$ is already quite high in many biological systems. To obtain some intuition for the numerical work, we analyzed four initial population states: (1) cell-I is initially OFF and everyone else is initially ON, (2) cell-I is initially OFF and everyone else is initially OFF, (3) cell-I is initially ON and everyone else is initially ON, and (4) cell-I is initially ON and everyone else is initially OFF. We computed phenotype diagrams for each of these four population-level states using the above algorithm. To obtain each phenotype diagram, we fixed the Hill coefficient $n$ to be one of the four aforementioned values, then simulated above algorithm for each value of $(K, S_{ON})$. Then we used a color to represent the phenotype score. Doing this for a wide range of values for $K$ and $S_{ON}$ resulted in a heat map. This is the phenotype diagram. We used a color spectrum that starts from a pure red (representing a phenotype score of 1) and ends in a pure blue (representing a phenotype score of 0). Thus a pure red represents an equilibrium value of $S_O$ that is close to $S_{ON}$ while a pure blue represents an equilibrium value of $S_O$ that is close to $S_{OFF} = 1$. An intermediate color such as green represents a case in which we obtained an intermediate value for the final value of $S_O$ in our iterative simulation. This means that we cannot say whether the cell is ON or OFF (Figs. S9 & S10).

As a consistency check, we note that for $n=40$, the simulations yield partial phenotype diagrams that closely resemble the phenotype diagram that we analytically computed for the cells with an infinite Hill coefficient (Fig. 3D). Note that our computed phenotype diagrams (Figs. S9 & S10) are partial phenotype diagrams (like Fig. 3D) because they represent what happens when we start with a particular initial population state $(C, \Omega)$. To get full phenotype diagrams (like Fig. 4C), we need to run our simulation for all possible population states $(C, \Omega)$, then super-impose all of them to get a single, full phenotype diagram. We have not done this here because the (partial) phenotype diagrams of the four limiting values of $(C, \Omega)$ already give us the boundary lines for the full phenotype diagram (i.e., the lines that separate the distinct phenotypes in Fig. 4C).

Moreover, we note that our simulations yield important differences between a population of $N$ cells with a finite Hill coefficient and a population of $N$ cells with an infinite Hill coefficient. This difference is starkest when all cells in the population are initially OFF (Figs. S9 & S10).

In the end, except for the appearance of intermediate states (e.g., green regions in Figs. S9 & S10), our work shows that the main qualitative features of the phenotype diagrams for cells with a finite Hill coefficient are essentially identical to the main features of the phenotype diagrams for the cells with an infinite Hill coefficient (Fig. 3D).

## 7. Calculation of the boundaries in the phenotype diagram for the basic population unit - Figure 4A

We choose cell-I to be at the center of the hexagon and consider $L < L_c$. It senses concentration $S_I$:

$$S_I = S_O + \sum_{j=1}^{6} S_{Oj} exp(-1/L), \qquad (47)$$

where we have used the same notations as in the main text. We can bin the 26 distinct population states (i.e., the $\Omega$'s) into seven equivalence classes: $\Omega_0$,..., $\Omega_6$. $\Omega_n$ denotes any population with n ON cells and 6-n OFF cells at the corners of the hexagon (n=0,...,6). Every population state $(C, \Omega)$ in a given class of population $(C, \Omega_n)$ has the same phenotype function $\phi_{C,\Omega_n}$. We denote $\phi_{0,\Omega_n}$ as $A_n(K, S_{ON}, L)$ and $\phi_{1,\Omega_n}$ as $D_n(K, S_{ON}, L)$. Then we have

$$A_n(K, S_{ON}, L) = S_{ON} - \frac{1}{ne^{-1/L}}K + \frac{1 + (6-n)e^{-1/L}}{ne^{-1/L}}, \tag{48}$$

and

$$D_n(K, S_{ON}, L) = S_{ON} - \frac{1}{1+ne^{-1/L}}K + \frac{(6-n)e^{-1/L}}{1+ne^{-1/L}} \tag{49}$$

In the case of $n = 0$, we have $A_0(K, S_{ON}, L) = -K + 1 + 6exp(-1/L)$ and $D_0(K, S_{ON}, L) = S_{ON} - K + 6exp(-1/L)$. Let us fix a value for the signaling length $L$. Then for each $n$, we find the relationship between $K$ and $S_{ON}$ that causes $A_n$=0 (i.e., the values of $(K, S_{ON})$ that cause $A_n$=0 or $D_n$=0 in the above equation yield straight lines) and another relationship between $K$ and $S_{ON}$ that causes $D_n$=0 (i.e., values of $(K, S_{ON})$ that cause $D_n$=0 in the above equation form a straight line). In the end, we obtain a total of fourteen lines that divide the plane spanned by $(K, S_{ON})$ into fifteen regions (Figure 4A: right panel). This forms the phenotype diagram of the basic unit. The activation region "$\widetilde{A_n}$" is bounded by the lines that correspond to $A_{n-1} = 0$ and $A_n = 0$. The deactivation region deactivation region "$\widetilde{D_n}$" is bounded by the lines corresponding to $D_n = 0$ and $D_{n+1} = 0$, (n=0...5).

## 8. Note on the periodic boundary condition used on calculating the signaling strength $f_N(L)$

Throughout our work, we used a periodic boundary condition that joins the edges of the two-dimensional lattice on which the cells are placed: The North edge is joined with the South edge, and the West edge is joined with the East edge). This leads to a sheet of cells forming a torus ("donut" shape). Thus our work treats populations of cells as a closed tissue. In this set up, any one cell sees the other cells the same way as any other cell would. Thus no cell is special in this closed tissue. For this reason, the "cell-I" that we introduced in our formalism can be any cell in the population. Moreover, the signaling strength function $f_N(L)$ is a purely geometric property and is the same value for every cell in the population.